\documentclass[fleqn,usenatbib]{mnras}

\usepackage{newtxtext,newtxmath}

\usepackage[T1]{fontenc}

\DeclareRobustCommand{\VAN}[3]{#2}
\let\VANthebibliography\thebibliography
\def\thebibliography{\DeclareRobustCommand{\VAN}[3]{##3}\VANthebibliography}

\usepackage{graphicx}	
\usepackage{amsmath}	
\usepackage{enumitem}   

\title[Deep learning in galaxy formation]{Predictions for the abundance and clustering of H$\alpha$ emitting galaxies.}

\author[M. S. Madar et al.]{
Makun S. Madar,$^{1}$\thanks{E-mail: makun.s.madar@durham.ac.uk}
Carlton M. Baugh,$^{1}$
and Difu. Shi$^{1}$
\\
$^{1}$Institute for Computational Cosmology, Department of Physics, Durham University, South Road, Durham, DH1 3LE, UK.\\
}

\date{Accepted XXX. Received YYY; in original form ZZZ}

\pubyear{2024}

\begin{document}
\label{firstpage}
\pagerange{\pageref{firstpage}--\pageref{lastpage}}
\maketitle

\begin{abstract}
We predict the surface density and clustering bias of H$\alpha$ emitting galaxies for the \textit{Euclid} and \textit{Nancy Grace Roman Space Telescope} redshift surveys using a new calibration of the \texttt{GALFORM} galaxy formation model.  We generate 3000 \texttt{GALFORM} models to train an ensemble of deep learning algorithms to create an emulator. We then use this emulator in a Markov Chain Monte Carlo (MCMC) parameter search of an eleven-dimensional parameter space, to find a best-fitting model to a calibration dataset that includes local luminosity function data, and, for the first time, higher redshift data, namely the number counts of H$\alpha$ emitters. We discover tensions when exploring fits for the observational data when applying a heuristic weighting scheme in the MCMC framework. We find improved fits to the H$\alpha$ number counts while maintaining appropriate predictions for the local universe luminosity function. For a flux limited \textit{Euclid}-like survey to a depth of $2\times10^{-16}~\textrm{erg}^{-1}~\textrm{s}^{-1}~\textrm{cm}^{-2}$ for sources in the redshift range $0.9<z<1.8$, we estimate 2962-4331 H$\alpha$ emission-line sources deg$^{-2}$. For a \textit{Nancy Grace Roman} survey, with a flux limit of $1\times10^{-16}~\textrm{erg}^{-1}~\textrm{s}^{-1}~\textrm{cm}^{-2}$ and a redshift range $1.0<z<2.0$, we predict 6786-10322 H$\alpha$ emission-line sources deg$^{-2}$.

\end{abstract}




\section{Introduction}
\label{sec:Introduction}

Forecasting the performance of cosmological surveys plays a central role in planning the survey strategy and evaluating how trade-offs in depth and solid angle will affect the science goals.  The wide field redshift surveys planned with \textit{Euclid} \citep{Laureijs2011, EuclidWide2022}  and the \textit{Nancy Grace Roman Space  Telescope} \citep{Spergel2015,Wang2022} will mainly sample H$\alpha$ emitters to map the cosmic large-scale structure. The figure-of-merit of cosmological probes that use galaxy clustering is dependent upon the number density and clustering strength of the galaxies being targeted \citep{Albrecht:2006}. This is still relevant for \textit{Euclid} post-launch, as the performance of the various detectors is assessed in situ and changes may be required to the survey strategy \citep{EuclidWide2022}. \textit{Roman} is due for launch in May 2027. 

There are two routes to making this characterisation of the redshift survey galaxies: exploiting existing studies of the target galaxy population to fit empirical models or using physically motivated models to predict the properties of the sample.
\citet{Pozzetti2016} attempted to describe the H$\alpha$ luminosity function (LF) estimates available at the time using empirical models. Three empirical models were fit to the H$\alpha$ LFs measured using the \textit{Hubble Space Telescope} (\textit{HST}) Wide Field Camera 3 (WFC3) Infrared Spectroscopic Parallels (WISP) \citep{Colbert2013}, Hi-Z Emission Line Survey (HiZELS) \citep{Geach2008, Sobral2009, Sobral2012, Sobral2013} and the HST Near Infrared Camera and Multi-Object Spectrometer (NICMOS) \citep{Shim2009}. 
The resulting simple functional forms for the H$\alpha$ LF   
can be integrated to obtain the number counts. The uncertainties were considerable, with the predicted surface density of H$\alpha$ emitters barely being constrained to within a factor of two.

Recently, with the addition of further space data, the situation has improved somewhat, and there have been renewed efforts to estimate the number of H$\alpha$ emitters that \textit{Euclid} and  \textit{Roman} are likely to observe \citep[e.g.][]{Colbert2013, Mehta2015, Valentino2017, Merson2018, Zhai2019, Wang2022, Zhai2021}.
\citet{Bagley2020} constructed a new data sample of line emitters from several \textit{HST} surveys and forecast the properties of H$\alpha$ (and [OIII]) emission line galaxies for future surveys. The results from \cite{Bagley2020} show a clear preference for the so-called `pessimistic' model 3 from \cite{Pozzetti2016}, which predicted the lowest surface density of emission-line galaxies (ELGs).  

 With a physical model, it is possible to predict the clustering of the galaxies as well as their abundance (see, for example, \citealt{Orsi2010, Merson2019}). \cite{Merson2018} used the \texttt{Galacticus}  semi-analytical model of galaxy formation \citep{Benson2012} to forecast the number density of H$\alpha$ emitters using a variety of dust attenuation models. \cite{Merson2018} predict 3900-4800 emitters per sq degree for the \textit{Euclid} selection. However, in this case, \texttt{Galacticus} was calibrated to reproduce a variety of observational constraints with particular emphasis on the local Universe, without any explicit reference to ELGs. This situation was rectified in \citet{Zhai2019}, in which \texttt{Galacticus} was recalibrated using a new N-body simulation simulation, the UNIT run \citep{Chuang2019} and different calibration data, which included the H$\alpha$ luminosity function from HiZELS \citep{Geach2008, Sobral2009, Sobral2013}.

Efficient calibration and exploration of galaxy formation models have been investigated in several papers, typically in two forms: a direct exploration of the model parameter space, running the full simulation for each set of parameters, and emulation or interpolation, in which the full calculation is mimicked by a cheaper process. Despite semi-analytical models (SAMs) being vastly cheaper to run than hydrodynamic simulations, direct exploration of their parameter space is still computationally expensive due to the large number of model evaluations required for an extensive search. 

Direct exploration examples include \citet{Kampakoglou2008}, who used Markov Chain Monte Carlo (MCMC) to calibrate a SAM to multiple datasets. MCMC was used again in \citet{Henriques2009} to calibrate their SAM, where they found that the choice of calibration dataset changed the values of the best-fitting parameters, pointing to deficiencies in their model. \citet{Lu2011, Lu2012} constrained the parameter space for their SAM using Bayesian inference to achieve acceptable fits to the $K$-band LF, this was expanded to include the HI mass function in \citet{Lu2014} (see also \citealt{Martindale:2017}). \citet{Ruiz2015} employed a stochastic technique called particle swarm optimization \citep{Kennedy1995} to calibrate the  \texttt{SAG} SAM \citep{Springel2001, Cora2006, Lagos2008, Padilla2014, Gargiulo2015} to the $K$-band LF. 

The second class of calibration involves building a statistical emulator of the SAM which can be evaluated much faster than running the full model, with the drawback of this being approximate by nature. \citet{Bower2010} and \citet{Vernon2010} constructed a Bayesian approximation technique (described in \citealt{Goldstein2007}) to the \texttt{GALFORM} model that can be rapidly evaluated at any point in parameter space to provide reasonable fits to the $K$- and $b_J$-band LFs. This work was extended in \citet{Benson2010} to explore how adaptable this reduced parameter space was to fit further observational datasets, and in \citet{Rodrigues2017} to calibrate \texttt{GALFORM} to the local galaxy stellar mass function.  \citet{Elliott2021} used a deep learning algorithm to emulate \texttt{GALFORM} across a range of output statistics. Elliott et~al. were able to run many simple MCMC chains to explore the parameter space and investigate how calibration to different datasets constrained the model parameters. The emulation method can cope with a high-dimension parameter space.

BuildingPlanck on \cite{Elliott2021}, we extend the calibration of \texttt{GALFORM} to forecast the number counts of H$\alpha$ emitters and their clustering bias.
We emulate \texttt{GALFORM} in the PLanck Millennium N-body simulation (\citealt{Baugh2019}; hereafter PMILL).
We use deep learning to build an emulator: this allows us to build flexible function approximators that can reveal non-linear relations within data without needing a pre-defined model. There have been many successful uses of deep learning in astronomy \citep[e.g.][]{Ravanbakhsh2016, Schmit2018, Perraudin2019, He2019, Cranmer2019, Ntampaka2019, Zhang2019, De2020}. We demonstrate the accuracy that can be achieved with deep learning when emulating \texttt{GALFORM} for the H$\alpha$ number counts. We can use a moderate number of training runs to achieve good accuracy when compared to other calibration methods outlined above. As a deep learning emulator can be evaluated much more rapidly than running \texttt{GALFORM}, we can run many MCMC chains to explore the parameter space and identify the range of parameters that fit the calibration datasets. We achieve this by minimizing the absolute error between the emulator output and the observational datasets, employing a heuristic weighting scheme to the various observational datasets. This automation of the model calibration allows us to exhaustively search the parameter space. 


The layout of this paper is as follows. We present the theoretical background in \S\ref{sec:Theory} and present the datasets relevant to this work. 
In \S\ref{sec:results} we present our results. In \S\ref{sec:rDataGen} we review the generation of the training and testing data, in \S\ref{sec:rEmulatorPerformance} we illustrate the predictive performance of the emulator, in \S\ref{sec:rParameter} we show the results of the model exploration and calibration and the results for the H$\alpha$ number counts and galaxy bias predictions. 
Finally, in \S\ref{sec:Discussion} we review the merits of our methods and outline potential future avenues. We assume a $\Lambda$CDM cosmology with $\Omega_M$=0.307 an $\Omega_{\Lambda}$=0.693, and $H_0$=67.77 kms$^{-1}$ Mpc$^{-1}$. 

\section{Galaxy Formation Model and Calibration Data}
\label{sec:Theory}

We give an overview of \texttt{GALFORM} (\S\ref{sec:GALFORM}), then in \S\ref{sec:DeepLearning} we give a brief review of deep learning and describe the emulator design, and in \S\ref{sec:Parameterfit} we discuss how we find best-fitting parameters using MCMC. In \S\ref{sec:Datasets} we outline the generation of training and testing data for the emulator and describe the observations used in the calibration. 

\subsection{\texttt{GALFORM}}
\label{sec:GALFORM}

\texttt{GALFORM} is a physically motivated semi-analytical galaxy formation model  \citep{Cole2000, Bower2006, Lacey2016}. \texttt{GALFORM} populates the DM haloes at the earliest branches of the halo merger tree with hot baryonic gas and models the main physical processes behind the formation and evolution of galaxies using a set of coupled differential equations, including (i) the collapse and merging of DM haloes, (ii) the shock-heating and radiative cooling of gas inside DM haloes, leading to the formation of galactic discs, (iii) quiescent star formation in galactic discs, (iv) feedback from SNe, active galactic nuclei (AGN), and photo-ionization of the inter-galactic medium, (v) chemical enrichment of stars and gas and (iv) dynamical friction driven by mergers of galaxies within DM haloes, forming spheroids and triggering starbursts. Full descriptions of these physical processes are given in  \citet{Lacey2016} (see also the reviews by \citealt{Baugh2006} and \citealt{Benson2010r}).

\texttt{GALFORM} distinguishes between central and satellite galaxies within their host dark matter halo, with some of the physical processes being affected by this designation. Central galaxies are placed at the centre of the most massive sub-halo and are the focus of all the gas that is undergoing cooling. Halo merger events choose the central galaxy of the main (most massive) progenitor halo as the central galaxy of the descendant halo with other galaxies becoming satellites. In the default gas cooling model (see \citealt{Font2008} for an alternative model), satellite galaxies are stripped of their hot gas as soon as they become satellites, hence quenching any further cooling and stopping any long-term star formation. Merger time scales are calculated based on the initial energy and angular momentum of the satellite's orbit (which is a random quantity), and the mass of the satellite and halo hosting both the satellite and central galaxy. It is expected that after this time the effects of dynamical friction will have caused the satellite to merge with the central galaxy. However, the merger time scale of a satellite is calculated every time a satellite's host halo merges into a sub-halo of a more massive halo \citep{Cole2000,Simha:2017}. 

Here we give an overview of the processes in \texttt{GALFORM} that are explored. The model parameters varied are listed in Table 1.

\subsubsection{Quiescent star formation in discs}
\label{sec:Quiescent}

The quiescent mode of star formation takes place in the disk following the accretion of cooled gas from the hot halo. The star formation rate (SFR) in the disk is calculated using the empirical law inferred from observations by \citet{Blitz2006} (as implemented in \texttt{GALFORM} by \citealt{Lagos2011}; see also \citealt{Fu2010} and \citealt{Popping2014} for the incorporation of similar schemes into other SAMS) which is based on observations of nearby star-forming disc galaxies. The SFR is assumed to be proportional to the mass of the molecular component of the gas in the disk $M_{\textrm{mol,disk}}$
\begin{equation}
    \psi_{\textrm{disk}}=\nu_{\textrm{SF}}M_{\textrm{mol,disk}},
    \label{eq: vsf}
\end{equation}
where $\nu_{\textrm{SF}}$ is the value of the SFR coefficient, which controls the rate of conversion of the molecular gas into stars in quiescent galaxy disks. This is an adjustable parameter set within the range inferred from observations by \citet{Bigiel2011}. The mass of molecular gas depends on the gas pressure in the mid-plane of the disk.

\subsubsection{Supernova feedback}
\label{sec:SN}

Supernovae (SNe; mainly Type II) eject gas from galaxies and their host dark matter halos. The model, therefore, assumes the rate of gas ejection due to supernova feedback is proportional to the instantaneous SFR $\psi$, with a mass loading factor that is dependent on the galaxy circular velocity, $V_{\textrm{c}}$, as a power law:
\begin{equation}
    \dot{M}_{\textrm{eject}} = \left( \frac{V_{\textrm{c}}}{V_{\textrm{SN}}} \right)^{-\gamma_{\textrm{SN}}} \psi,
    \label{eq:Masseject}
\end{equation}
where $\gamma_{\textrm{SN}}$ and $V_{\textrm{SN}}$ are adjustable parameters. We can further split the $V_{\textrm{SN}}$ term into $V_{\textrm{SN, disk}}$ and $V_{\textrm{SN, burst}}$ to distinguish the feedback contributions in quiescent star formation in disks from star formation in bursts. Most studies have assumed that these velocity normalisation parameters are equal (e.g. \citealt{Gonzalez2014} and  \citealt{Lacey2016}). However, recent versions of the model have relaxed this restriction \citep[e.g.][]{Benson2010, Elliott2021}.  

Gas ejected from the galaxy due to SN feedback is assumed to gather in a reservoir beyond the virial radius of the host dark matter halo. The gas gradually returns to the hot gas reservoir within the virial radius at a rate of 
\begin{equation}
    \dot{M}_{\textrm{return}} = \alpha_{\textrm{ret}}\frac{M_{\textrm{res}}}{\tau_{\textrm{dyn, halo}}},
    \label{eq:Massreturn}
\end{equation}
where $\tau_{\textrm{dyn, halo}}$ is the halo dynamical time, $M_{\textrm{res}}$ is the mass of the reservoir beyond the virial radius, and $\alpha_{\textrm{ret}}$ is a free parameter. 

\subsubsection{Galaxy mergers}
\label{sec:merge}

It is assumed when galaxies merge there may be a burst of star formation and destruction of the galactic disks. To define the type of merger we set two thresholds, $f_{\textrm{ellip}}$ and $f_{\textrm{burst}}$. These thresholds are compared to the baryonic masses of the central galaxy, $M_{\textrm{b,cen}}$, and the merging satellite galaxy, $M_{\textrm{b,sat}}$ through the ratio  $M_{\textrm{b,sat}}$/$M_{\textrm{b,cen}}$. When $M_{\textrm{b,sat}}$/$M_{\textrm{b,cen}}\geq f_{\textrm{ellip}}$, the merger is classified as a \textit{major} merger. After a major merger, the disk component of the primary galaxy is destroyed and forms a spheroid. We assume the cold gas in the disk is used up in a burst of star formation which also adds stars to the spheroid. The case for which  $M_{\textrm{b,sat}}$/$M_{\textrm{b,cen}}$<$f_{\textrm{ellip}}$ is a \textit{minor} merger. Following a minor merger, the disk survives. For the cold gas in the disk to be consumed in a starburst after a minor merger, we require  $M_{\textrm{b,sat}}$/$M_{\textrm{b,cen}} \geq f_{\textrm{burst}}$. Both $f_{\textrm{ellip}}$ and $f_{\textrm{burst}}$ are free parameters. We use the prescription of \cite{Simha:2017} to compute the time for a galaxy merger to take place.

\subsubsection{Disk instabilities}
\label{sec:Disk}

Disk instabilities can trigger star formation. When a galaxy is dominated by rotational motion the disk is unstable to bar formation through sufficient self-gravitation. We assume that disks are dynamically unstable to bar formation if the following condition is met \citep{Efstathiou1982} 
\begin{equation}
    F_{\textrm{disk}} \equiv \frac{V_{\textrm{c}}(r_{\textrm{disk}})}{(1.68GM_{\textrm{disk}}/r_{\textrm{disk}})^{1/2}} < F_{\textrm{stab}},
    \label{eq:Diskinstable}
\end{equation}
where $M_{\textrm{disk}}$ is the total disk mass and  $r_{\textrm{disk}}$ is the disk half-mass radius. $F_{\textrm{disk}}$ describes the contribution of disk self-gravity to its circular velocity, with larger values equating to lower self-gravity and greater disk stability. Predictions of $F_{\textrm{disk}}$ vary depending on the method; \citet{Efstathiou1982} found $F_{\textrm{disk}}\approx$ 1.1 for a suite of exponential stellar disk models, while \citet{Christodoulou1994} found $F_{\textrm{disk}}\approx$ 0.9 for gaseous disks. For a completely self-gravitating stellar disk,  $F_{\textrm{disk}}=$ 0.61.  $F_{\textrm{stab}}$ is a model parameter. 

If the disk instability condition $F_{\textrm{disk}}$<$F_{\textrm{stab}}$ is met we assume the disk forms a bar which evolves into a spheroid \citep{Combes1990, Debattista2006}. We assume that an unstable disk is disrupted by bar instabilities on a sub-resolution timescale thus all the mass is instantly transferred to the spheroid and any gas present is used in a burst of star formation. 

\subsubsection{Starbursts}
\label{sec:starburst}

We assume the rate at which bursts of star formation form stars in a spheroid is  
\begin{equation}
    \psi_{\textrm{burst}} = \nu_{\textrm{SF,burst}}M_{\textrm{cold,burst}}=\frac{M_{\textrm{cold,burst}}}{\tau_{\textrm{* burst}}},
    \label{eq:psiburst}
\end{equation}
where the timescale $\tau_{\textrm{* burst}}$ is
\begin{equation}
    \tau_{\textrm{* burst}} = \textrm{max}[f_{\textrm{dyn}}\tau_{\textrm{dyn,bulge}}, \tau_{\textrm{* burst,min}}].
    \label{eq:tauburst}
\end{equation}
The bulge dynamical time is defined in terms of the half-mass radius and circular velocity of the bulge, $\tau_{\textrm{dyn,bulge}}$=$r_{\textrm{bulge}}/V_{\textrm{c}}(r_{\textrm{bulge}})$. We treat $\tau_{\textrm{* burst,min}}$ as a parameter, but fix $f_{\textrm{dyn}}$=20 \citep{Lacey2016}. 

\subsubsection{Stellar Initial Mass Function and Stellar Population Synthesis}

We assume that quiescent star formation in galactic disks produced stars with a solar neighbourhood stellar initial mass function (IMF). Bursts of star formation (triggered by mergers or dynamically unstable disks) produce stars with a top-heavy IMF, with a power law slope of $x=1$ (see \citealt{Lacey2016}). We use the stellar population synthesis models of \cite{Maraston:2005}.

\subsubsection{SMBH growth and AGN feedback}
\label{sec:SMBH}

Supermassive black holes (SMBH) can inject energy into the halo gas disrupting gas cooling. Multiple instances can lead to black hole growth; hot halo accretion, BH-BH mergers, and starbursts \citep{Bower2006, Fanidakis2011, Griffin2019}. In a starburst, mass accreted onto the SMBH is a constant fraction of the mass of stars formed, $f_{\textrm{SMBH}}$, where $f_{\textrm{SMBH}}$ is a parameter. AGN heating of the hot gas halo is assumed to occur if two conditions are met: (1) the gas halo is in quasi-hydrostatic equilibrium, i.e.  
\begin{equation}
    \tau_{\textrm{cool}}/\tau_{\textrm{ff}} > 1/\alpha_{\textrm{cool}}, 
    \label{eq:alphacool}
\end{equation}
where $\tau_{\textrm{cool}}$ is the gas cooling time, $\tau_{\textrm{ff}}$ is the free-fall time, and $\alpha_{\textrm{cool}}$ is a parameter; (2) the AGN power required to balance the radiative cooling luminosity is less than  $f_{\textrm{Edd}}$ times the SMBH  Eddington luminosity.

\subsubsection{Emission lines}

The star formation histories predicted by \texttt{GALFORM} are convolved with a simple stellar population model, which gives the light emitted as a function of age by a population of stars produced with a given stellar initial mass function and metallicity, building up a composite spectral energy distribution (SED) for each galaxy (see the review by \citealt{Conroy2013}). 
This information is used to compute the number of ionising photons, which, along with the metallicity of the cold gas in the interstellar medium is combined with an HII region model to compute the luminosity of the emission lines (see \citealt{Cole2000} and \citealt{Baugh2022} for more extensive descriptions of the emission line models). For some predictions we combine the H$\alpha$ and N[III] line luminosities, as these lines are close together in wavelength and will not be fully resolved by the surveys we 
consider.

Dust is assumed to be mixed with the stars in two forms: in clouds and a diffuse component \citep{Granato2000}. Dust properties are assumed and combined with the predicted scalelengths of the disk and bulge allowing the optical depth and attenuation of the starlight to be calculated as a function of wavelength. The emission lines are assumed to have the same attenuation due to dust as the stellar continuum at the same wavelength. 

\subsection{Deep learning emulator}
\label{sec:DeepLearning}


We now describe the construction of an efficient emulator of \texttt{GALFORM} using  \texttt{tensorflow} \citep{Abadi2016}. This is a supervised learning problem (also known as associative learning) in which the emulator is trained by providing it with inputs and matching outputs. We define the input vector $\mathbf{x}$ to represent a set of \texttt{GALFORM} model parameters and predict an output vector $\mathbf{y}$, which consists of the binned statistical properties of the resulting synthetic galaxy population, for example, the galaxy luminosity function. The emulator aims to map the input vector $\textbf{x}$ to the output vector $\mathbf{y}$ via an unknown function $\hat{f}(.)$ which replaces running the full  \texttt{GALFORM} model at a fraction of the computational cost. The emulator allows us to thoroughly search a multi-dimensional model parameter space. The problem is one of regression where the outputs are binned floats rather than the probabilities that might be found in classification problems. 


We use an artificial neural network to emulate \texttt{GALFORM}. 
The first layer of the multi-layer network  is the input layer with a size equal to the number of entries or components in $\mathbf{x}$. In our case, this is the number of \texttt{GALFORM} input parameters used to make the predictions, with one neuron per feature. Note that these input parameters are the subset that is being varied; the full parameter space of the model is larger than the 11 parameters that we vary here, but the other parameters are held fixed (for the full list of parameters see Table 1 in \citealt{Lacey2016}). The final layer is the output layer with one neuron per prediction value. Here the number of output neurons is the total number of bins across all of the chosen statistics. The middle layers of the network are known as hidden layers. The neurons in these layers extract features for mapping an input to an output and the network is trained by evaluating the hidden layer neurons using labelled examples, i.e. with the output from runs of \texttt{GALFORM}. Networks with multiple hidden layers are known as deep learning networks. The connections between each neuron have an associated weight, $w$, and each neuron has a bias, $\theta$. A network learns by adjusting these weights and biases from exposure to the training examples according to some learning rule. Each neuron is a simple mathematical function taking a vector of inputs and calculating an output. The $i$-th neuron in the $j$-th layer contains a vector of adjustable weights $\mathbf{w}_{ij}$ and an adjustable bias $\theta_{ij}$. The vector $\mathbf{w}_{ij}$ contains all the weights linking a neuron $i$ to each neuron in the previous layer, $j-1$. The data flow from the input to output neurons is strictly passed forward and every neuron in each layer is connected to every neuron in the following layer in what is known as a fully connected network. Note there are no connections \textit{within} a layer. The total input of neuron $i$ in layer $j$ is a function of the outputs from each neuron in layer $j-1$,  $\mathbf{y}_{j-1}$, the neuron vector weights $\mathbf{w}_{ij}$, and the bias of the neuron $\theta_{ij}$. An activation function $\mathcal{F}(.)$ takes the total input of the neuron to produce an output,
\begin{equation}
    y_{ij} = \mathcal{F}(\mathbf{w}_{ij} \cdot \mathbf{y}_{j-1} + \theta_{ij}).
    \label{eq: neuron_output}
\end{equation}

The activation function is often a non-decreasing function of the total input of the neuron, introducing non-linearity to the network and allowing for complex representations and functions to develop, which is not possible with a simple linear input-output model. The activation function transforms the output value of the neuron to within certain limits, modified based on the application of the model. If unrestricted by the activation function, the outputs of neurons can explode in magnitude in deeper networks. Generally, some sort of non-linear threshold function is used, such as a sigmoid or hyperbolic tangent function. The outputs of the neurons, $y_{ij}$, are passed to the following layers of neurons, and so on, until the final layer is reached. The output from the final layer is the network prediction $\mathbf{y}$ from input $\mathbf{x}$. An activation function is still applied to the final layer but this is usually a linear function in the case of regression. 

To adjust the weights assigned to hidden neurons we use the back-propagation learning rule \citep{Rumelhart1986}. During training the predictions from the output layer are compared to the true values and the error between these two are back-propagated from the output layer to the hidden layers and their weights are adjusted accordingly to minimize an error function. Following \citet{Elliott2021} we minimize the mean absolute error function (MAE) between the emulator predictions of the \texttt{GALFORM} outputs and the true outputs

\begin{equation}
    \textrm{MAE} =\frac{1}{n}\sum_{k=1}^{n}|\hat{\mathbf{y}}_k-\mathbf{y}_k|,
    \label{eq: MAE}
\end{equation}
where $\hat{\mathbf{y}}_k$ is the emulator prediction for the $k$-th sample out of $n$ and $\mathbf{y}_k$ are the values computed by \texttt{GALFORM} for the same parameter values. The MAE is also known as the loss function and reveals how badly (or how well) the network is performing. 

The neural network is trained iteratively over many epochs. One epoch is equivalent to the network cycling through every sample in the training set once; the number of training epochs is a user choice. An optimizer algorithm is used to change the weights and biases of the neural network by seeking minima on the error surface, often via a form of gradient descent. The optimizer also specifies the size of steps taken during the gradient descent towards the local minima, known as the learning rate. At the end of each epoch, the adjusted model is tested on a validation sample, which is a subset of the data that has not been used during the training to ensure the model is generalisable to completely unseen data. The number of training epochs is fixed by plotting the MAE against the epoch; this curve flattens off after some number of training epochs so that the precise choice of the number of epochs used is not important once this flat part of the MAE curve has been reached (see e.g. Fig.~\ref{fig:Activation}). 

The final network is tested on a hold-out set of samples to carry out a performance analysis on completely unseen data (\S~2.4.1).

\subsubsection{Inputs and outputs}
\label{sec:INOUT}

\begin{table}
 \caption{The \texttt{GALFORM} parameter space investigated assuming a uniform range for each parameter. See \S~\ref{sec:GALFORM} for an explanation of how each process is modelled and the equations which involve each parameter. The first column gives the parameter name (and units if relevant), the second column gives the range over which the parameter is allowed to vary and the third column lists the process to which the parameter relates.
 }
 \label{tab:paramrange}
 \begin{tabular*}{\columnwidth}{@{}l@{\hspace*{20pt}}l@{\hspace*{20pt}}l@{}}
  \hline
  Parameter & Range & Process\\
  \hline
   $\nu_{\textrm{SF}}$ [Gyr$^{-1}$] & 0.1 - 4.0 & Quiescent star formation\\ 
   $V_{\textrm{SN, disk}}$ [kms$^{-1}$] & 10 - 800 & SN feedback\\
   $V_{\textrm{SN, burst}}$ [kms$^{-1}$] & 10 - 800 & SN feedback\\
   $\gamma_{\textrm{SN}}$ & 1.0 - 4.0& SN feedback\\
   $\alpha_{\textrm{ret}}$ & 0.2 - 3.0 & SN feedback\\
   $\textrm{F}_{\textrm{stab}}$ & 0.5 - 1.2 & Disk instability\\
   $f_{\textrm{ellip}}$ & 0.2 - 0.5 & Galaxy mergers\\
   $f_{\textrm{burst}}$ & 0.01 - 0.3 & Galaxy mergers\\
   $\tau_{\textrm{* burst,min}}$ [Gyr] & 0.01 - 0.2 & Starbursts\\ 
   $f_{\textrm{SMBH}}$ & 0.001 - 0.05 & SMBH growth\\
   $\alpha_{\textrm{cool}}$ & 0.0 - 4.0 & AGN feedback\\

  \hline
  \label{tab: range}
 \end{tabular*}
\end{table}

We aim to develop an emulator to map an input vector $\textbf{x}$, which is the subset of \texttt{GALFORM} parameters that are allowed to vary, onto an output vector $\mathbf{y}$,  corresponding to the statistical galaxy properties we wish to predict. Our choice of the input parameters that are allowed to vary is made through physical intuition and guidance from previous analyses (see \S~\ref{sec:Datasets}). 
These parameters and their ranges are shown in Table~\ref{tab: range}. We tune the emulator to predict three statistical properties calculated from the output of \texttt{GALFORM} to calibrate a model to make accurate predictions for \textit{Euclid} and \textit{Roman}: these are the redshift distribution of H$\alpha$ emitters between $0.69 \le z \le 2$, and the local luminosity functions in the $r$ and $K$-bands (see \S~\ref{sec:CalibrationData} for more information about these datasets). Each dataset is weighted equally in the metric when the emulator is being constructed. 

\subsubsection{Network architecture}
\label{sec:Network}

\begin{figure}
    \centering
    \includegraphics[trim=0.9cm 0.9cm 0.9cm 0.9cm,width=1.03\columnwidth]{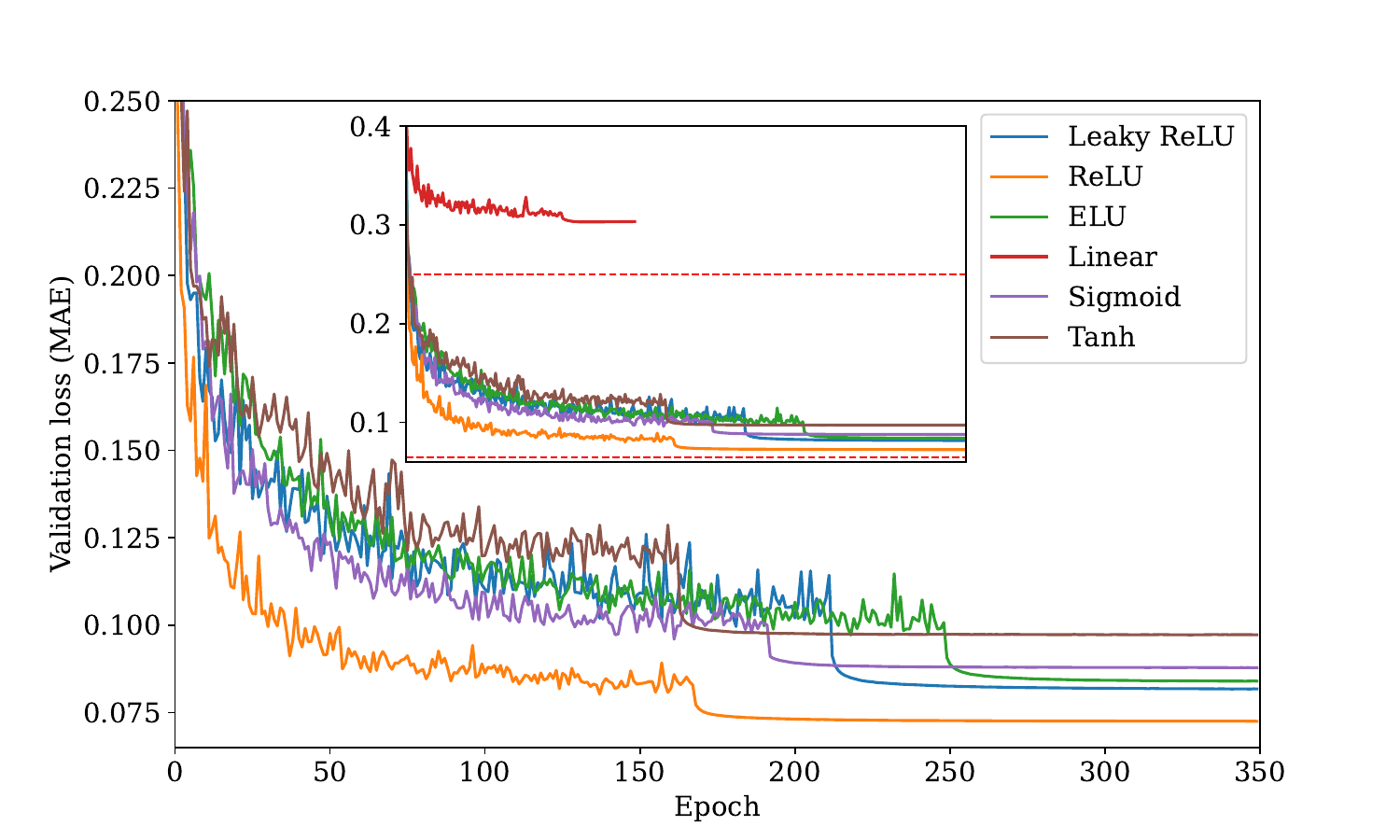}
    \caption{Testing the choice of activation function in the network. The MAE loss on the validation dataset is plotted against the training epoch. A different colour is used for each choice of activation unit, as indicated by the key. Each network has the same architecture of 2 hidden layers, with 512 nodes and a linear output activation function. We display a zoomed-out inset to show the poor loss attained with a linear activation function. The sudden drop in loss value exhibited in all cases, when the curves also appear to become smoother, is due to the fine-tuning stage of training (see text for further details).}
    \label{fig:Activation}
\end{figure}

\begin{figure}
    \centering
    \includegraphics[trim=0.9cm 0.9cm 0.9cm 1.2cm,width=1.03\columnwidth]{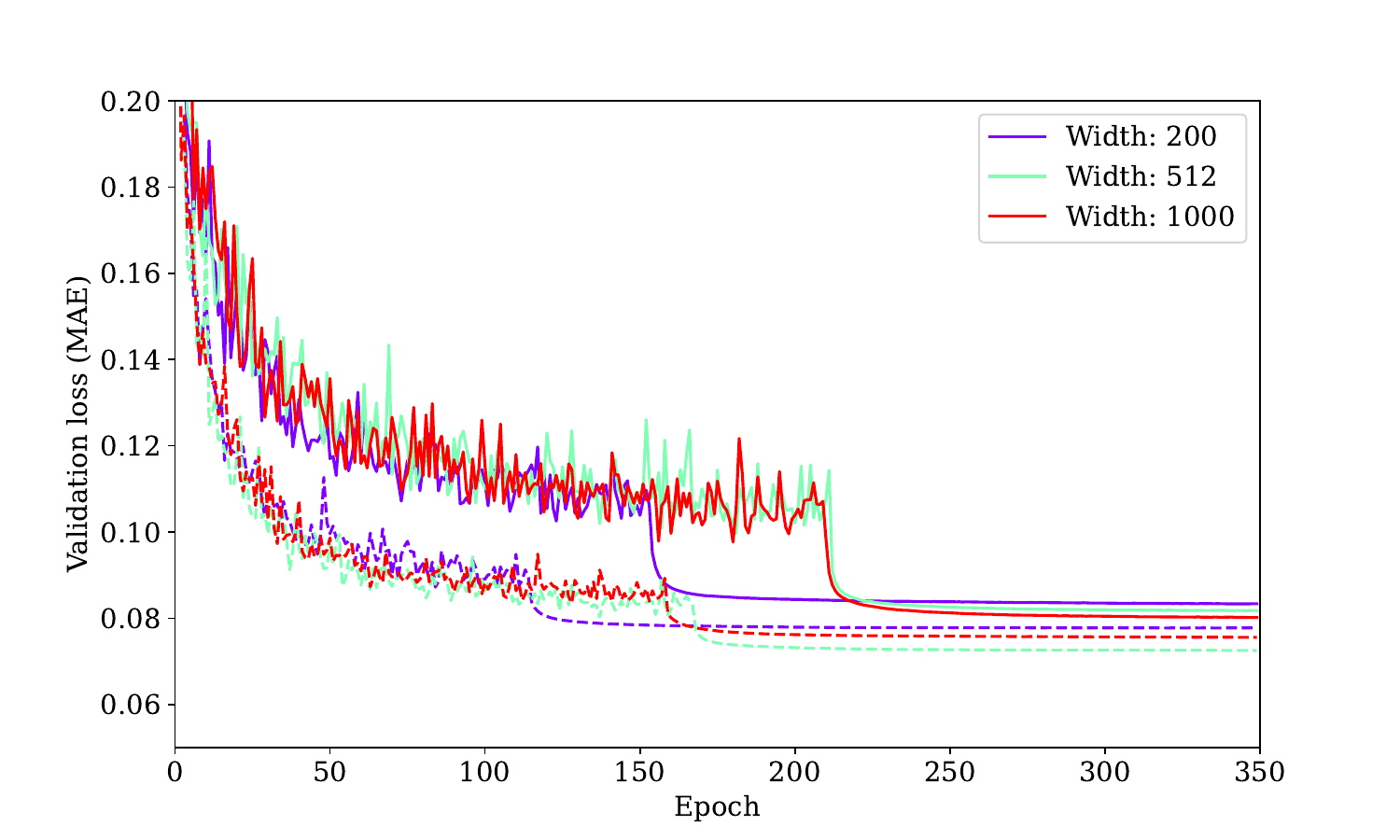}
    \caption{Measuring the MAE loss on the validation dataset during training, when altering the hidden layer widths of our network for two activation functions, ReLU (dashed) and LReLU (solid). Each network has two hidden layers and a linear output activation function. There are no significant benefits to increasing the width of our network beyond 512 neurons per layer.}
    \label{fig:Width}
\end{figure}

\begin{figure}
    \centering
    \includegraphics[trim=0.9cm 0.9cm 0.9cm 1.2cm,width=1.03\columnwidth]{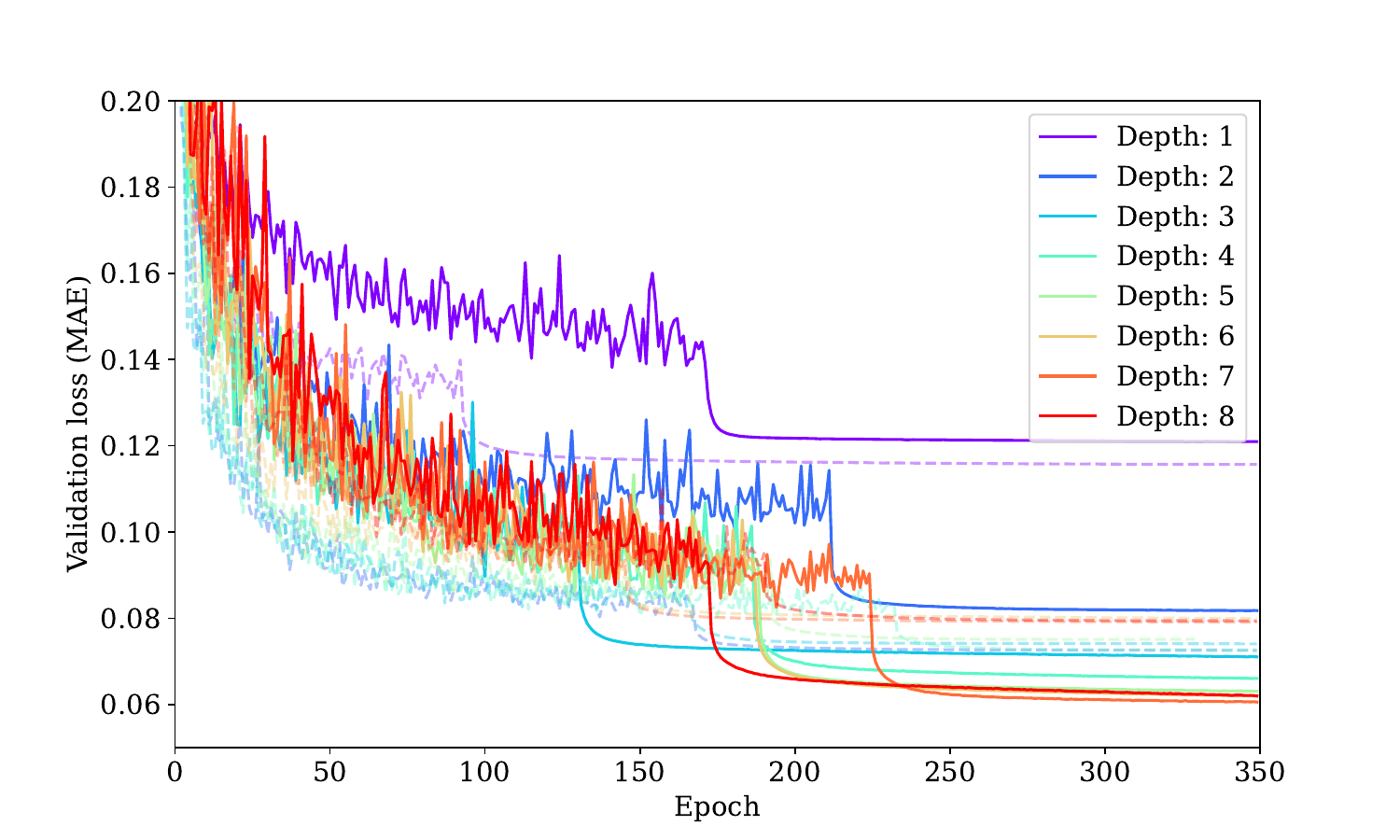}
    \caption{MAE validation loss when modifying the number of hidden layers in the network, with different colours indicating different numbers of layers, as shown by the key. We keep the width of the network fixed at 512 and show results for two activation functions, LReLU (solid) and ReLU (dashed). The LReLU function has greater potential for improvement than the ReLU networks. An increase in depth improves the performance of our network up to a depth of five or six layers. Beyond this, there is only a modest improvement in the MAE at the expense of an increase in the computational cost.}
    \label{fig:Depth}
\end{figure}

The neural network architecture was determined by testing individual hyperparameter configurations. Taking inspiration from \citet{Elliott2021}, we start with an architecture with two hidden layers, each containing 512 neurons with the sigmoid activation function on hidden layers, and linear activations on the output layer. Here, we test modifying the choice of activation function on hidden layers, the width of the network (the number of neurons per layer), and the depth of the network (the number of hidden layers). For the output layer, the linear activation function is consistently used, which is suitable given that the emulator is essentially a regression model. All networks are trained with the same dataset. We track the MAE against the validation dataset at each epoch during training and show the results in Figs.~\ref{fig:Activation}, \ref{fig:Width} and \ref{fig:Depth}. We note there is a caveat with these tests due to the stochastic nature of training a neural network; an identical network architecture trained on identical training data can display a small variability in its final validation score, so we take this into account when deciding on the final network. 

Starting from the architecture used in \citet{Elliott2021}, we modify the activation functions, testing a linear function, Logistic Sigmoid, Tanh, Rectified Linear Unit (ReLU) \citep{Nair2010, Sun2015}, Leaky ReLU (LReLU) \citep{Maas2013, Xu2015}, and Exponential Linear Unit (ELU) \citep{Clevert2015}, with the results displayed in Fig.~\ref{fig:Activation} (for a full review of the many activation functions available see \citealt{Dubey2022}). We found that modifying the activation function to a type of rectifier unit was the best option. 

Next, we test both the ReLU and LReLU activation functions while modifying the width of our network but keeping the number of hidden layers at two. We consider 200, 512, and 1000 neurons per hidden layer. We want to see if there is a positive trend in terms of a reduction in the MAE when increasing the number of neurons per layer. In Fig.~\ref{fig:Width} we plot the results from both the LReLU (solid line) and the ReLU (dotted) network activation functions. There are training speed benefits to using a thinner network: the percentage increase in training speed for the network to reach epoch 350, between the thinnest network (width 200) and the widest network (width 1000) is $\sim190\%$ for either activation function. We see that for both cases the 200-width network does not perform as well as the wider networks. However, there is no clear gain in performance to support increasing the width beyond 512 neurons. Therefore we will use hidden layer widths of 512 to optimise the performance and training speed. 

Finally, we test the depth of the network, that is, the number of hidden layers our network contains. Once again we train two identical networks, one with a LReLU activation function, and the other with the ReLU activation function, shown in Fig.~\ref{fig:Depth} as solid and dashed lines respectively. An interesting observation is the improvement seen with the LReLU network when more layers are included. In Fig.~\ref{fig:Activation} we saw the ReLU activation function performs best when two hidden layers were used, but as the number of hidden layers increases the performance increases with the LReLU network putting it ahead of all of the ReLU. Furthermore, we do see performance gains when increasing the number of hidden layers up to a certain number when they start to converge on a minimum MAE loss. We find, that for both activation functions, once there are five hidden layers, there are no further significant gains in network performance when more layers are used. Computational speed again is a factor here, with the training time needed for a network with eight hidden layers being 217 per cent longer than for one with a single hidden layer. Our final network architecture, based on the results presented here, is six hidden layers, each with 512 neurons and LReLU activation functions. 

Having made this choice, we explain in more detail the difference between a ReLU and a Leaky ReLU (LeLU). A Leaky ReLU builds from the original ReLU by modifying the handling of negative input values. The ReLU returns an output of zero for a negative input,
\begin{equation}
    \mathcal{F}(s_{ij}) = \textrm{max}(0, s_{ij}), 
    \label{eq:ReLU2a}
\end{equation}
whereas a Leaky ReLU assigns a non-zero slope on the negative end,
\begin{equation}
    \mathcal{F}(s_{ij}) = \textrm{max}(\alpha s_{ij}, s_{ij}).
    \label{eq:LReLU2a}
\end{equation}
In Eqn.~\ref{eq:LReLU2a} $\alpha$ is a hyperparameter generally set to 0.01, and $s_{ij}$ is the total input to neuron $i$ in the $j$th layer. The Leaky ReLU solves the `dying' ReLU problem \citep{Lu2019}, where a standard ReLU can become inactive and output zero for any input value. In this case, it can never recover and can lead to network regions becoming `inactive'. We find using a Leaky ReLU instead of ReLU improves the MAE performance during training on the validation set, reducing the average MAE loss by $\sim~29$ per cent. 

We use the Adaptive Momentum Estimation (Adam) optimiser which is a popular momentum-based gradient descent optimization algorithm \citep{Kingma2014, Reddi2019} and set the learning rate to 0.005. We add the AMSGRAD variation \citep{Tran2019} which aims to improve the performance of Adam around the minima on the error surface using a stochastic method, which evaluates the weights after every mini-batch iteration (mini-batches are small subsets of the whole training set). At the end of each epoch, we save the model weights if the performance on the validation set has improved (as measured by Eqn.~\ref{eq: MAE}) and continue training until there is no improvement for 30 epochs. Then the learning rate is reduced to $10^{-5}$ for a fine-tuning stage with the RMSprop optimizer \citep{Tieleman2012}, allowing us to take small steps towards the minimum of the error surface. RMSprop uses stochastic gradient descent and assumes the error surface is a quadratic bowl. This method boosts the performance of the emulator as we can descend into fine local minima, and we measure improvements to our network by tracking the MAE of the validation samples throughout training. We see evidence of this in Figs~\ref{fig:Activation}, \ref{fig:Width} and \ref{fig:Depth} where the MAE rapidly drops when the network transitions into fine-tuning training mode. 

\subsubsection{Ensembling}
\label{sec:Ensemble}

Before training, the weights of a network are initialized according to some distribution, often random. We use an initialiser described in \citet{Glorot2010}. Due to the stochastic nature of the training process training a single network is insufficient since the error surface is likely to contain many local minima and one network is unlikely to traverse enough of the weight space to find the best possible mapping. Over-fitting is also a potential problem due to the large number of parameters especially as more layers are added. One solution to these issues is ensembling multiple network predictions \citep{Opitz1999, Sagi2018, Ganaie2022}.
This involves training several identical networks with different weight initializations and shuffling the validation and training sets for each model in the ensemble so the models are distributed from input to output. This should allow for a more robust final prediction. We average over the predictions of each model to negate any over- or under-fitting to different features of the data. 

Using this method, we train 5 separate networks, each with the same model architecture. The final emulator prediction is the average of the predictions from the ensemble of models.  There is scope in the future to improve on this method via a method called stacking \citep{Wolpert1992} where the ensemble networks themselves are the inputs to a single network with generalizes the outputs for improved results. However, this works best where the ensemble networks are varied and provide different information, such as different architectures or combining different types of machine learning algorithms. 

\subsection{Parameter fitting}
\label{sec:Parameterfit}

We use the emulator for inference on target datasets; that is, fitting the model to given datasets. We employ a MCMC sampler to compare the generated models against the observed datasets with the goal of sampling from a set of parameters that produces the models that best fit the observables. The Metropolis-Hastings algorithm \citep{Robert2004} is a common and simple method of executing an MCMC, generating serially correlated draws from a sequence of probability distributions, eventually converging to a given target distribution. The means of convergence comes from the minimization of the absolute error between the emulator output and the observational constraints. We note that as we are minimizing the absolute error between multiple data sets i.e. the discrepancy between the model prediction and the data, we do not take into account their associated (measurement)z errors. We wish to weight certain datasets over others to allow us to investigate the effect of requiring better fits to some datasets and how this affects the reproduction of other datasets, as well as seeing how the optimal parameter choices change as a result. We therefore introduce a modified version of the MAE (introduced in Eqn~\ref{eq: MAE}) which includes a vector of heuristic weights, $\mathbf{W}$, to vary the contribution of the residuals from constraint $i$ to the total error,
\begin{equation}
    \textrm{MAE}^{\textrm{obs}}(\mathbf{y},\hat{\mathbf{y}}) = \frac{1}{n^{\textrm{obs}}}\sum_{i=1}^{n^{\textrm{obs}}}\frac{\mathbf{W}_i}{n^{\textrm{obs}}_{i}}\frac{|\mathbf{y}_i - \hat{\mathbf{y}}_i|}{\boldsymbol{\sigma}_i},
    \label{eq: MAEmod}
\end{equation}
where $\mathbf{\hat{y}}_i$ is the predicted value of the $i$-th observable constraint, and $\mathbf{y}_i$ is the corresponding observable value across $n^{\textrm{obs}}_{i}$ datapoints. Due to $\mathbf{\hat{y}}_i$ and  $\mathbf{y}_i$ being vector quantities, the modulus represents the L1 norm. $\boldsymbol{\sigma}_i$ is a vector of errors corresponding to $\mathbf{y}_i$. We sum over the $n^{\textrm{obs}}$ observable constraints. The different observational datasets contain different numbers of data points, therefore we divide the weighted absolute error of the $i$-th dataset by the number of data points, $n^{\textrm{obs}}_{i}$, for equal contribution to the mean error result. In later sections, we refer to Eqn~\ref{eq: MAEmod} as the mean absolute error (MAE). 

The Metropolis-Hastings procedure for updating a Markov Chain compares the likelihoods from the current parameter location or state to a proposed (new) state. Assuming uniform priors throughout, each chain is initialized on a random point in the parameter space which is assigned as the \textit{current} state, $\mathbf{x}$. Then we sample a proposed state, $\mathbf{x'}$, from independent Laplacian proposal distributions about $\mathbf{x}$, $\mathcal{L}(\mathbf{x'}|\boldsymbol{\mu}, \mathbf{b})=\left(1/2\mathbf{b}\right)\textrm{exp}(-|\mathbf{x'}-\boldsymbol{\mu}|/\mathbf{b})$ where $\boldsymbol{\mu}$=$\mathbf{x}$ and the scale parameter vector $\mathbf{b}$ is set as 1/20th of the parameter ranges given in Table \ref{tab: range}. The proposed state must satisfy the condition that the proposal lies within the defined parameter bounds given in Table~\ref{tab:paramrange}. We decide whether the proposed state is accepted or not by measuring the likelihood improvement of emulator predictions to the observational data from the current to the proposed state using a Laplacian likelihood with scale parameter $b_{\textrm{obs}}$ = 0.005. Taking the ratio of likelihoods at states $\mathbf{x'}$ and $\mathbf{x}$ gives the \textit{acceptance ratio}, $\alpha$, 
\begin{equation}
    \alpha =  \frac{\mathcal{L}(f_*(\mathbf{x'})|\boldsymbol{\mathbf{y}},b_{\textrm{obs}})}{\mathcal{L}(f_*(\mathbf{x})|\mathbf{y},b_{\textrm{obs}})},
\end{equation} 
where $\mathbf{y}$ is the vector of observables and $f_*(\cdot)$ is the emulator.  We could use a ratio of errors as an acceptance ratio in our MCMC, however, doing so may not align with the principles of Bayesian inference and so could have implications for the accuracy and efficiency of our algorithm. The likelihood is often used in Bayesian inference due to its probabilistic interpretation, as it provides a measure of how well the model explains the observed data given a set of parameters. The acceptance ratio is compared to an acceptance criterion, \textit{u}, which is a random uniform number \textit{u}$\in$~[0, 1]; a proposed state is \textit{accepted} if $\alpha \geq$ \textit{u}, in which case $\mathbf{x}=\mathbf{x'}$ and the next sample is drawn from a Laplacian centred on the new state, or a proposed state is \textit{rejected} if $\alpha$ < \textit{u} for which case we sample again from the original Laplacian centred on $\mathbf{x}$. Using this method, if the error between the emulator predictions and the observables reduces when moving from state $\mathbf{x}$ to $\mathbf{x'}$ the sample is always accepted, else we accept the proposed state $\mathbf{x'}$ with a probability $\alpha$. We expect the density of accepted samples to trace the regions in the parameter space which give the best fits to the data. At the start of the chain, there will be a burn-in phase as the accepted samples tend towards local maxima in the parameter space so we discard the first half of accepted samples. Testing multiple chain lengths we find chains converge to local MAE minima (given by Eqn.~\ref{eq: MAEmod}) within the first 5,000 samples and so we choose the chain length as 7,500 (after discarding the burn-in phase).


\subsection{Datasets}
\label{sec:Datasets}

Our decision as to which \texttt{GALFORM} input parameters to vary comes from a combination of physical motivation and choices informed by previous analyses (mainly \citealt{Elliott2021}). We differ from their parameter choices by focusing more on the contribution from quiescent galaxies and less on galaxies experiencing a starburst. For a typical H$\alpha$ galaxy survey, we find burst galaxies only affect the extremely bright end of the luminosity function and have little impact on the overall number counts (see for example the predictions from \citealt{Lacey:2011} for the UV LF, which is also sensitive to recent star formation). Close to the \textit{Euclid} and \textit{Roman} flux limits, quiescent galaxies are dominant. Burst galaxies do, however, dominate the high flux tail of the H$\alpha$ emitter counts, but this is not important for the overall clustering measurement. 

 
\subsubsection{Training and testing data}
\label{sec:TrainTestData}
We use a supervised machine-learning method to emulate running a computationally expensive model, \texttt{GALFORM}. Training the emulator requires running the full model. Generally, the more samples used during training, the better the predictions should be. 

Whereas \cite{Elliott2021} ran \texttt{GALFORM} at a single output redshift to make predictions for their  calibration data, the computational cost per model is much higher in our case as we need to compute the redshift distribution of H$\alpha$ emitters. This is due to the structure of the \texttt{GALFORM} code; running for $N$ output redshifts effectively increases the run time by a factor of $N$. 
One option to produce predictions for the redshift distribution of H$\alpha$ emitters would be to generate a lightcone catalogue \citep[e.g.][]{Merson2013}. For the PMILL simulation, for the range of redshifts of interest for \textit{Euclid} and \textit{Roman}, this would require running 135 output redshifts. Instead, we can run \texttt{GALFORM} for a much smaller, select number of output redshifts, taken from the target range. For each output, we construct the LF of H$\alpha$ emitters. We then use this information to compute the redshift distribution of H$\alpha$ emitters, interpolating between the luminosity functions computed at the output redshifts. Another computational saving that can be made is to run \texttt{GALFORM} for a fraction of the available dark matter halo merger histories. We experimented with using different numbers of output redshifts and different fractions of the merger histories to compute the redshift distribution of H$\alpha$ emitters, comparing the answers to a full lightcone calculation. 
The calculation converges with five PMILL redshift snapshots between redshifts $0.69 \le z \le 2.00$ using $\sim0.5\%$ of the available halo merger histories. We also produce output at $z=0$ to compare to the local calibration data. 


Model parameters were generated via Latin hypercube sampling for efficient and smooth coverage  \citep[as described in][]{Loh1996,Bower2010}. The parameter ranges are given in Table~\ref{tab: range}.
The Latin hypercube sampler generated 3000 sets of the 11 parameters,  resulting in 3000 \texttt{GALFORM} outputs, each with an associated H$\alpha$ redshift distribution and $z=0$ $K$- and $r$-band LFs. The \texttt{GALFORM} inputs and outputs formed the input-output vector pairs, (\textbf{x$_i$}, \textbf{y$_i$}) for the deep learning emulator, where \textbf{x$_i$} is the $i$th set of model parameters and \textbf{y$_i$} is the corresponding output vector of the redshift distribution and LFs. We separate the samples randomly into three sets:  training, validation, and holdout sets. The training and validation sets are used during the training of the emulator, and the hold-out set is kept separate to evaluate performance on unseen data. The ratio of training samples to hold-out test samples was 29:1 and for each network trained, 20\% of the training samples were randomly chosen as the validation data.

\subsubsection{Calibration and comparison datasets}
\label{sec:CalibrationData}

 Traditionally, \texttt{GALFORM} has been calibrated mostly using local data, as these have been the measurements with the smallest errors (see \citealt{Lacey2016}). We continue this trend by using the $r$ and $K-$band LFs measured from the Galaxy And Mass Assembly (GAMA) survey \citep{Driver2012}; here, these data replace the older $b_{\textrm{J}}$ and $K$-band LFs used to calibrate \texttt{GALFORM}. This choice has the advantage that the same team has done the data reduction and made the assumptions about the $k$- and  evolutionary corrections. We also use the H$\alpha$ redshift distribution measured by  \cite{Bagley2020}. Using calibration datasets at different redshifts greatly reduces the volume of the viable parameter space.  

The full list of calibration and comparison datasets is as follows: 
\begin{enumerate}[wide, labelwidth=!, label=(\roman*)]
    \item For the H$\alpha$ redshift distribution, we calibrate the emulator to the redshift distribution from \citet{Bagley2020}. They used measurements from two slitless spectroscopic WFC3-IR datasets, 3D-HST+AGHAST and the WISP survey \citep{Atek2010} to construct a \textit{Euclid}-like sample. They detect the combined H$\alpha$+[NII]-emission from  galaxies in the redshift range $0.9\le z\le1.6$ with total line flux brighter than  $\ge2\times10^{-16} \textrm{erg s}^{-1} \textrm{cm}^{-2}$.
    \item For the $z=0$ $K$-band and $r-$band LFs, we take data from \citet{Driver2012} who used the GAMA dataset to construct the low-redshift ($z<0.1$) galaxy luminosity functions in multiple bands. 
\end{enumerate}

We also compare our best-fitting models to the previous local LF calibration data (the $K-$ band LF measured by \citealt{Cole2001} and the $b_{\textrm{J}}$ measured by \citealt{Norberg2002}). This is to see if models using the new local calibration datasets still give good fits to the old calibration data; this is an indirect way of seeing (through a model) if these observational LFs are consistent with one another.

\section{Results}
\label{sec:results}

\subsection{\texttt{GALFORM} runs for training and testing}
\label{sec:rDataGen}

We start with the \citet{Lacey2016} model and replace the parameters highlighted in Table~\ref{tab: range}, using the 3000 combinations generated by the Latin hypercube sampler. For each model, we run \texttt{GALFORM} at six redshift snapshots $z=0,  0.69, 0.90, 1.14, 1.60, 2$. These were selected to cover the redshift range probed by the \citet{Bagley2020} H$\alpha$ redshift distribution and the local LFs.

\subsection{Emulator performance}
\label{sec:rEmulatorPerformance}


The development of the architecture for the emulator is described in \S\ref{sec:DeepLearning}. During the architecture training phase, we only ran our networks up to 350 training epochs. For the final network, we chose not to limit the number of epochs but instead included an early stopping clause which stopped and saved the network at its lowest training MAE validation loss score. On average the networks gave their lowest loss score between 500 and 700 epochs. We also follow  \citet{Elliott2021} by ensembling  networks of the same architecture, averaging over the outputs to produce a final result. We tested ensembles of five and ten networks and found little to no improvement in emulator  performance against the hold-out set. Note that going from one network to an ensemble of five gives roughly a ten per cent reduction in the MAE. Furthermore, the more networks to be averaged over, the greater the computational time which becomes important as we run an MCMC across a substantial number of walkers each with around 15\,000 steps. Therefore, the emulator consists of five equal architecture networks (described in \S~\ref{sec:DeepLearning}). We want to evaluate the ability of the emulator to output accurate \texttt{GALFORM} predictions at new points in the parameter space. The set of 3000  \texttt{GALFORM} outputs was split up with 96.67 per cent of the outputs used for training our emulator as described in \S~\ref{sec:DeepLearning} (equating to 2900 parameter combinations) and the remaining 3.33 per cent (100 parameter combinations) being used as unseen outputs for testing purposes (hold-out set). This split maximises the number of training samples and provides an appropriate range of unseen test samples to evaluate the network. When training each network, we randomly split the 2900 parameter output combinations into a training set and a validation set with 20 per cent going towards validation (580 parameter combinations). For each network trained in the emulator ensemble, the training and validation sets were shuffled. 

\begin{figure*}
    \centering
    \includegraphics[trim=0.5cm 0.95cm 0 0 , width=0.99\textwidth]{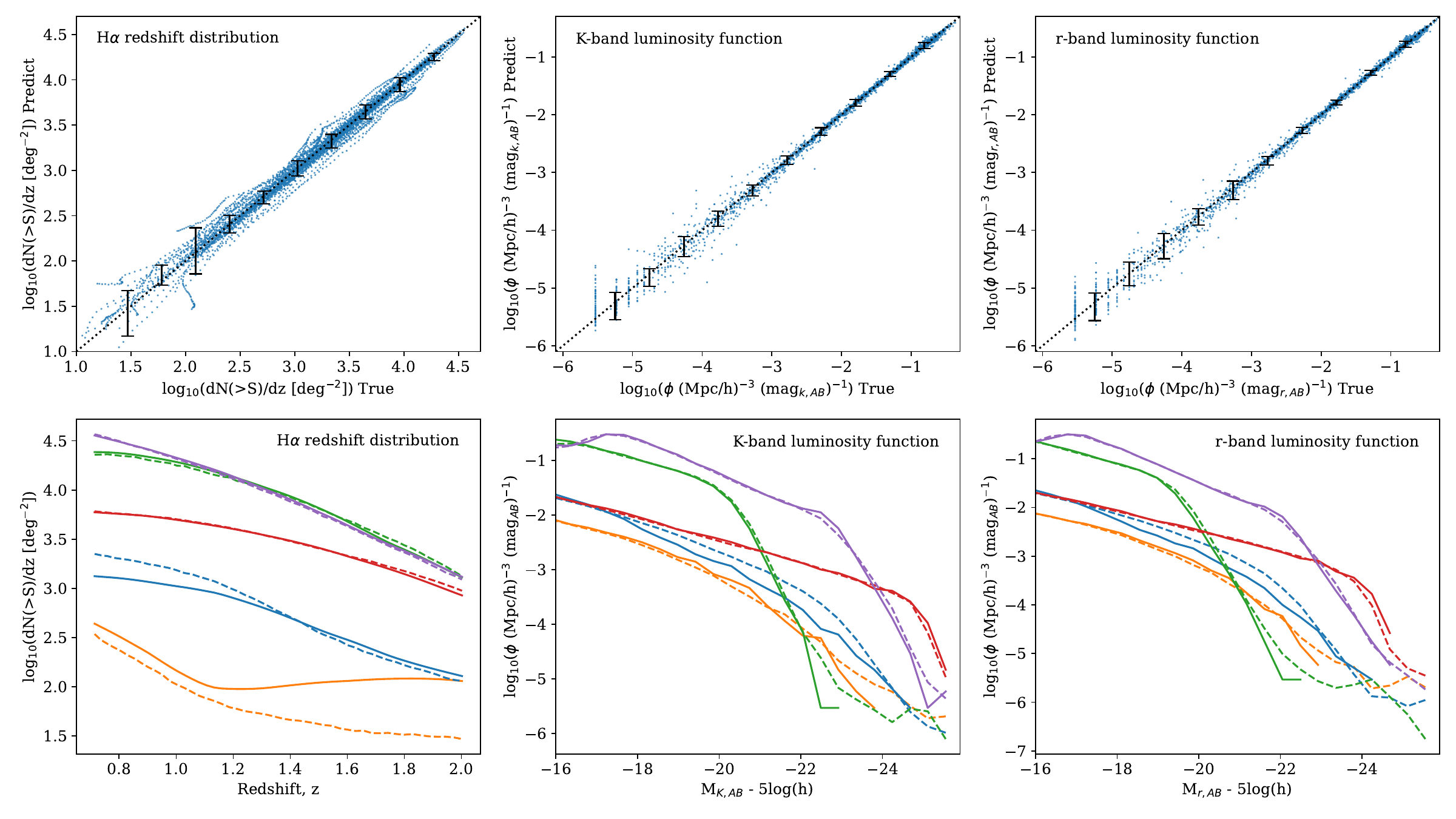}
    \caption{Emulator performance across the three calibration statistics computed with the holdout parameter sets. The top row shows the emulator output ($y$-axis) against the true \texttt{GALFORM} output ($x$-axis). Black error bars indicate the 10-90th percentile range of the residuals. The bottom row shows a draw of emulator outputs (dashed lines) and true \texttt{GALFORM} outputs (solid lines) for selected parameter sets. In these panels, different colours denote different parameter sets. }
    \label{fig:NNpredvtrue}
\end{figure*}
In the upper panels of Fig.~\ref{fig:NNpredvtrue} we show the emulator predictions against the hold-out set outputs from the corresponding full \texttt{GALFORM} runs. A perfect emulator would follow the $y=x$ line (dotted) with no scatter. In general, we see the emulator following a tight relation to the diagonal across the three statistics, indicating that the emulator is accurately predicting \texttt{GALFORM} output for the holdout set parameters, without any significant biases and a reasonably small scatter. Out of the three statistics, the redshift distribution predictions appear to have a greater uncertainty than the $K$ and $r$-band luminosity functions. However, this is largely an artefact of the redshift distribution predictions spanning a smaller dynamic range than the other statistics, so this scatter plot is `zoomed-in' compared to the others (covering just over 4.5 decades in scale as opposed to six decades in the other panels). In the lower panels of Fig.~\ref{fig:NNpredvtrue} we show the performance of the emulator across the three statistics on a sample of the holdout set parameters, plotting the emulator outputs as dashed lines and the true \texttt{GALFORM} outputs as solid lines. The parameter samples drawn from the holdout set were chosen to reflect the range of emulator performances, including parameters that the emulator most struggled with for each statistic. Each colour across the three panels is the same combination of parameters from the holdout set. The luminosity function plots display the ability of the emulator to predict beyond the resolution of \texttt{GALFORM} when the true model was generated with a finite sample of merger histories from the simulation, which can result in some luminosity bins being empty at the bright end. The lower panel of Fig.~\ref{fig:NNpredvtrue} reveals some sources of inaccuracies in the predictions, particularly the H$\alpha$ redshift distribution, which is more prone to exhibiting noisy behaviour for some choices of parameters, for example, the low redshift distribution (orange line) is poorly predicted. The error bars for the redshift distribution predictions are fairly even across the redshift bins and this is reflected in the lower panel plots where the majority of predictions follow the shape of the true \texttt{GALFORM} output but with varying degrees of offset. The main source of errors for the luminosity function predictions is seen at low values of $\phi$. We do see that at the bright end of the luminosity function plots the predictions can become noisy but the overall shape is well captured.  

The majority of emulator predictions for the redshift distribution are reasonably close to the \texttt{GALFORM} predictions, but we do come across cases with substantial discrepancies between the true and predicted outputs (as exhibited by the orange line in the bottom row of Fig.~\ref{fig:NNpredvtrue}). We see far fewer cases like this within the holdout set of poor predictions of the true \texttt{GALFORM} outputs when it comes to both luminosity functions, with the largest discrepancy seen in the blue parameter set. These poor predictions are usually indications that the training data did not contain sufficient examples of this behaviour as these examples appear to be extreme cases of the output and so are less common. The emulator constructs a function $f_*(\cdot)$ by fitting it to the training examples, where $f_*(\cdot)$ can interpolate between the points in the parameter space. However, the interpolation is less reliable in the sparser regions of the space, such as at the extremities of our parameter bounds.

We can see that at the bright ends of the $K$- and $r$-band LFs in Fig.~\ref{fig:NNpredvtrue}, the emulator tends to slightly over-predict the \texttt{GALFORM} output. This is a consequence of using a small fraction of the available merger histories (~0.6\% of the total), which leads to noisy predictions at low galaxy number densities, and, as seen in  Fig.~\ref{fig:NNpredvtrue}, cut-offs at different luminosities for different choices of parameters. The emulator outputs a fixed number of bins, therefore during training, we omit any luminosity bins which contain zero galaxies when computing the loss. This leads to the emulator having fewer brighter luminosity bins to fit which are biased towards having higher values of $\phi$ in these brighter bins. This causes more cases of over-prediction at these luminosities. This problem is minor since the \citet{Driver2012} luminosity function data does not sample $\phi$ to very low number densities. These issues could be resolved by evaluating \texttt{GALFORM} using a larger fraction of the available merger histories, although this would be more expensive computationally with little gain. 

\begin{figure*}
    \centering
    \includegraphics[trim=0.35cm 1.0cm 0 0.5cm, width=0.99\textwidth]{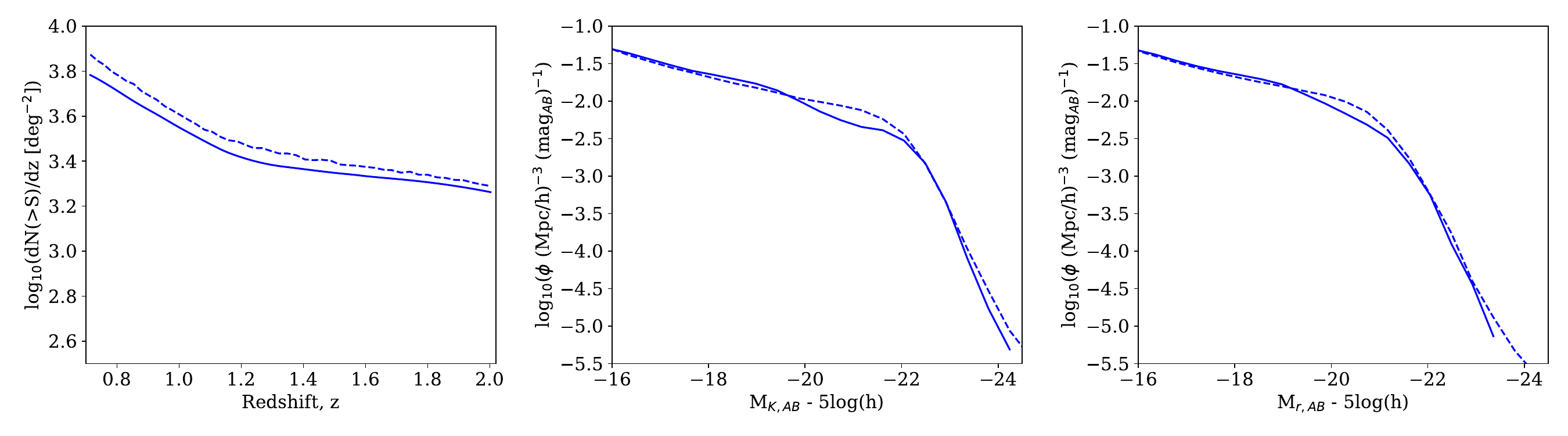}
    \caption{Emulator predictions (dashed lines) using the \citet{Lacey2016} \texttt{GALFORM} parameters compared with the true \texttt{GALFORM} outputs (solid lines). We predict the H$\alpha$ redshift distribution (left), and the $z=0$ $K$- (middle) and $r$-band (right) luminosity functions.}
    \label{fig:laceypred}
\end{figure*}

We also evaluate the performance of the emulator against the \citet{Lacey2016} \texttt{GALFORM} model in Fig.~\ref{fig:laceypred}.
We see an overall good fit to the true model, with the emulator redshift distribution overpredicting the true \texttt{GALFORM} model by a small amount. This matches our findings of the emulator performance on the holdout set above. For the redshift distribution, the emulator can still accurately identify the shape of the true model. The emulator does well at matching the true \texttt{GALFORM} model for the local LFs, with the only deviation seen around the break at magnitudes $\sim$-22 for the $K$-band and $\sim$-21 for the $r$-band. The emulator is unable to recreate the dipped features around these magnitudes which indicates a deficiency of these types of parameters within our training set. The possible changes we could make to the training set of the emulator that we highlighted before would improve our predictions against the \citet{Lacey2016} parameter set.  

\subsubsection{Performance and training set size}
\label{sec:rTrainingsize}

To find how  the emulator 
 performance depends on the number of full \texttt{GALFORM} calculations, we train the emulator with 900, 1900 and 2900 samples of parameters (in each case split with 20 per cent of the samples going towards validation). The emulators consist of an ensemble of five identical networks each trained on the same (shuffled) training and validation sets.  Performance is evaluated on the same 100 holdout parameter samples. 
The emulator shows a clear reduction in the MAE with an increasing number of training samples. Using an ensemble of networks results in a near-constant improvement in performance of almost 12 per cent compared to using a single network: however, this effect saturates after 5 networks.
    \label{fig:TrainingMAE}


\subsection{Parameter fitting on the calibration data - model optimisation}
\label{sec:rParameter}

\begin{figure}
    \centering
    \includegraphics[trim=0.85cm 1.cm 0 0,width=1.01\columnwidth]{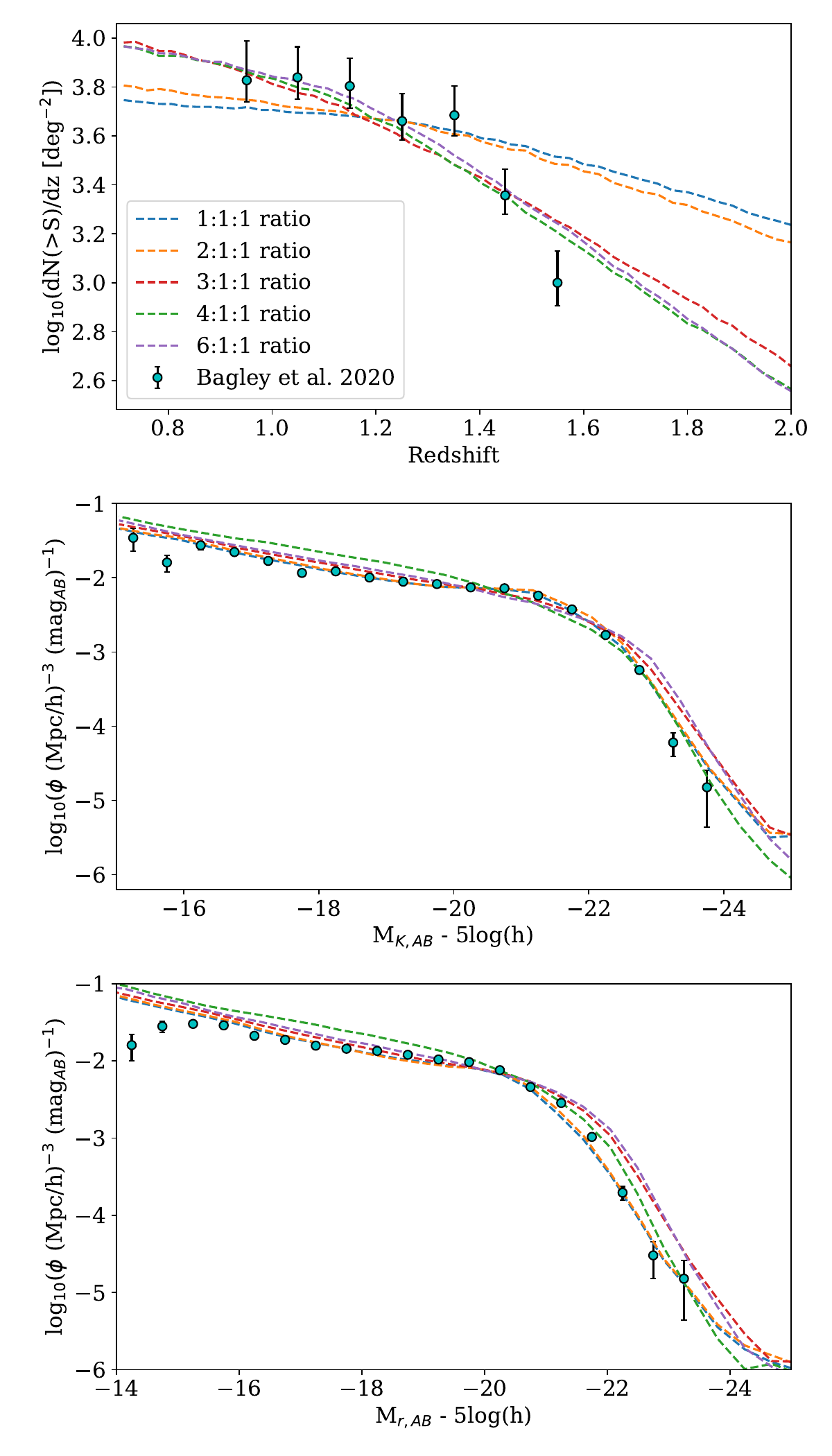}
    \caption{Best MCMC fits for five different weighting schemes (as indicated by the key in the top panel), on increasing the weight applied to the H$\alpha$ redshift distribution (first number in the line label) to display the tensions between the constraints (the other two numbers show the weights applied to the local LFs). We show a redshift distribution weight value $W$ of one (blue), two (orange), three (red), four (green) and six (purple), plotted with the \citet{Bagley2020} H$\alpha$ redshift distribution (top) and \citet{Driver2012} $z=0$ $K$- (middle) and $r$-band  luminosity functions (bottom).}
    \label{fig:weightingtest}
\end{figure}

We apply the methods described in \S~\ref{sec:Parameterfit} to calibrate the model to the datasets introduced in \S\ref{sec:CalibrationData}. We begin by investigating the tensions between the three statistics by adjusting the weights applied to the residuals between our emulator prediction and each dataset (given by Eqn.~\ref{eq: MAEmod}) and then performing an MCMC parameter search to see how the best-fitting parameter choices respond. In Fig.~\ref{fig:weightingtest} we show the emulator predictions for the best-fitting parameters found from five MCMC chains using different weighting schemes. To make accurate predictions for \textit{Euclid} and \textit{Roman} we need to fit the H$\alpha$ redshift distribution data from \citet{Bagley2020}. However, to reduce the overall model parameter space, it is important to constrain the model to reproduce the local luminosity functions. Hence we need to find a balance of fits between the two. When the weighting to the H$\alpha$ redshift distribution data is low, for example, a weighting of one or two (blue and orange lines in Fig.~\ref{fig:weightingtest} respectively), we see a poor accuracy reproduction of the H$\alpha$ redshift distribution data and strong performance regarding the luminosity functions, particularly around the break. As the redshift distribution weighting increases, we notice increasing deviation at the bright- and faint ends of the luminosity functions, but an improved fit to the redshift distribution data, with the predicted distributions being within the error bounds of the observations. Applying a  weight of four to the redshift distribution (green line) still allows us to recover the LF break at $L^*$ and stays just as close to the high redshift data points in the redshift distribution as a weight of six (purple line). The spread across the LFs for the different weightings is surprisingly low given that the spread in the redshift distribution fits is large in comparison. This could indicate that there are multiple regions in the parameter space that can fit these models, according to the emulator. This likely arises from the error of the emulator outputs, particularly for the redshift distribution predictions. It is worth noting that these parameter fits come from a small number of MCMC chains: we expect to see improvements in the best-fitting parameters when we evaluate 100 MCMC chains.


We set $W_i=4$ for the redshift distribution constraint, and apply unit weight to both $K$- and $r$-band LF constraints.
With the weighting scheme for the three statistics fixed, we recalibrate \texttt{GALFORM} across the three constraints to estimate the best-fitting parameters. We run 100 MCMC chains with our emulator, each with 7\,500 steps after the burn-in phase (which itself is 7\,500 steps). The residual of each sample is computed using the emulator and the weighted MAE function. The minimum MAE obtained for each chain lies in the range $\sim0.25-0.28$. As we have seen in \S\ref{sec:rEmulatorPerformance}, our emulator outputs have an associated error, so we can not confidently discern which parameter sets give the best fit to the observational data with the emulator alone. Hence, we evaluate the parameters that gave the lowest MAE value from each of the 100 MCMC chains with \texttt{GALFORM}.

\begin{figure*}
    \centering
    \includegraphics[trim=0 0.8cm 0 0, width=0.7\textwidth]{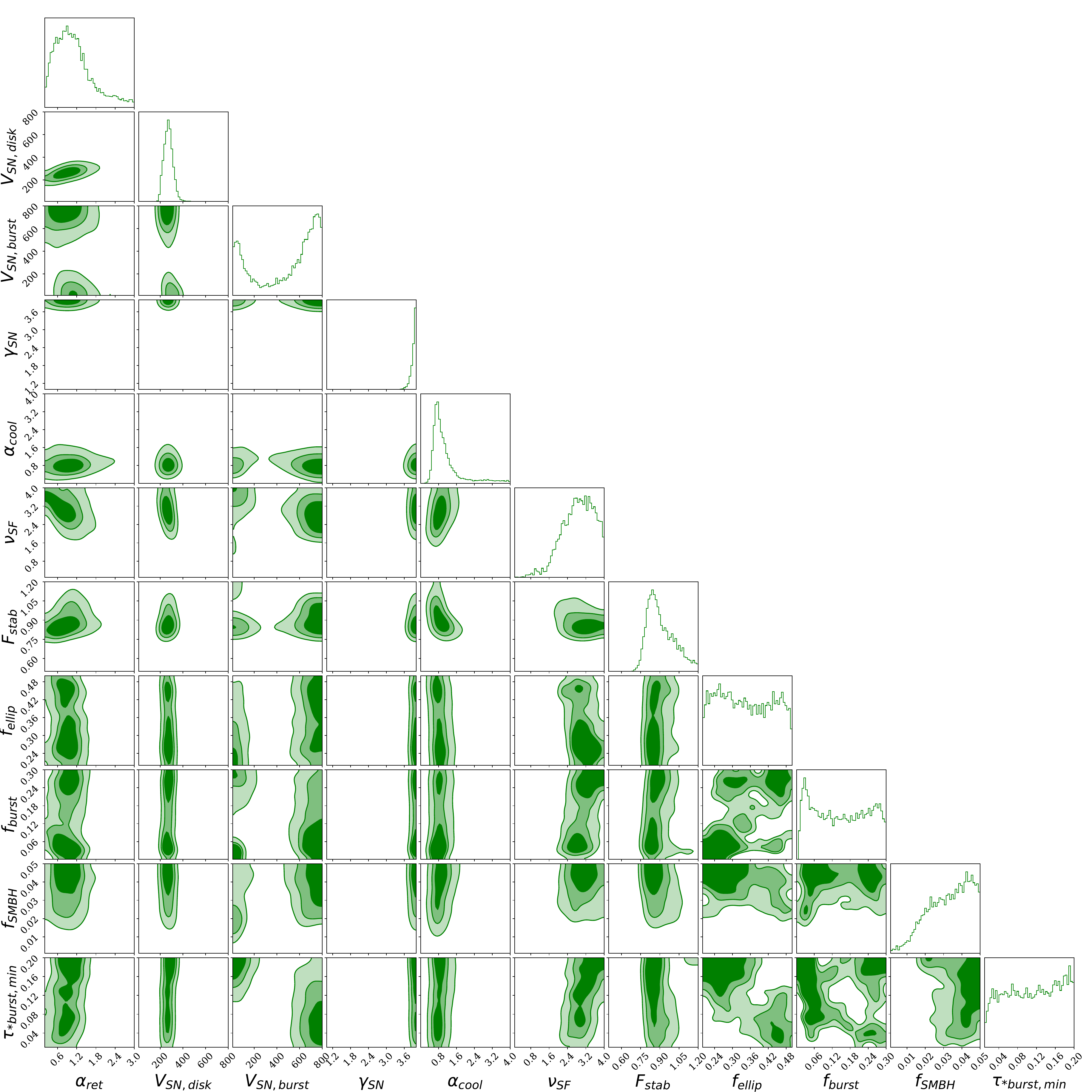}
    \caption{Accepted samples from 100 MCMC chains for fits to the H$\alpha$ redshift distribution, $K$- and $r$-band LFs. The first 50\% of samples were discarded to allow for burn-in. The histograms show the marginalised distribution of the parameters. The ranges on each axis are the same as those quoted in Table~\ref{tab: range}. The shading corresponds to the density of chain steps, with darker colours corresponding to more densely sampled regions. The darkest regions correspond to the 25th percentile and the lighter regions to the 50th and 75th percentiles.}
    \label{fig:Corner}
\end{figure*}

In Fig.~\ref{fig:Corner} we illustrate the regions in the parameter space sampled by the MCMC chains. The shaded regions show the accepted samples from our chains, each 7500 steps long after discarding the burn-in. The shading indicates the density of the accepted samples, with the darker regions corresponding to the more favoured regions of the parameter space. 
Also shown in Fig.~\ref{fig:Corner} are 1D histograms of the density of accepted samples. For some parameters, a reasonably large range of parameter values results in acceptable fits to the constraints. However, when plotted in one or two dimensions the space appears widely sampled, on moving to a higher dimension the acceptable regions are reduced significantly. This is the effect of the high dimensionality of the parameter space, as described in \citet{Bower2010}. We see that to fit the three statistics using the weighting scheme described, the fits prefer high values of $\gamma_{\textrm{SN}}\sim4$ possibly beyond the sampling parameter boundary. We have the option to extend our parameter space, but doing so will probe parameters beyond the space used to train the emulator. This could result in more uncertain predictions. Furthermore, we do not want to extend our parameter ranges to unphysical choices for the processes being modelled. We also observe a bimodal distribution for the $V_{\textrm{SN, burst}}$ parameter which tends towards the lower and upper boundaries of our parameter range at $\sim10$kms$^{-1}$ and $\sim 800$kms$^{-1}$ respectively. In contrast, the parameters $f_{\textrm{ellip}}$, $f_{\textrm{burst}}$, and $\tau_{\textrm{*burst,min}}$ are weakly constrained showing almost uniform sampling, whereas the parameters that contribute to the SN and AGN feedback are more tightly bound.

\begin{figure*}
    \centering
    \includegraphics[trim=0.5cm 1.5cm 0 0,width=0.99\textwidth]{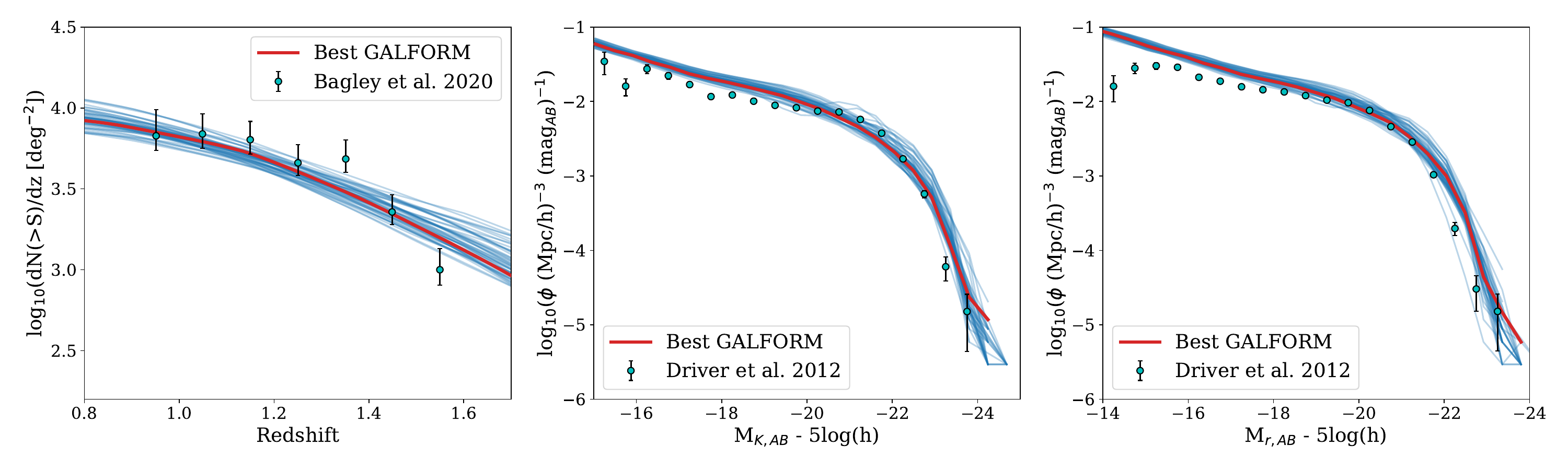}
    \caption{The \texttt{GALFORM} evaluations of the best-fitting parameters found with 100 MCMC chains, each 7,500 samples in length, using the constraint weightings described in the text. Here we plot a sample of the best 50 runs, as measured by weighted MAE (Eqn.~\ref{eq: MAEmod}). The red line indicates the parameter set with the lowest weighted MAE. The remaining 49 runs are plotted in blue. The data described in \S\ref{sec:CalibrationData} is shown in cyan.}
    \label{fig:50BestMCMC}
\end{figure*}

\begin{figure*}
    \centering
    \includegraphics[trim=0.5cm 1.5cm 0 1cm, width=0.99\textwidth]{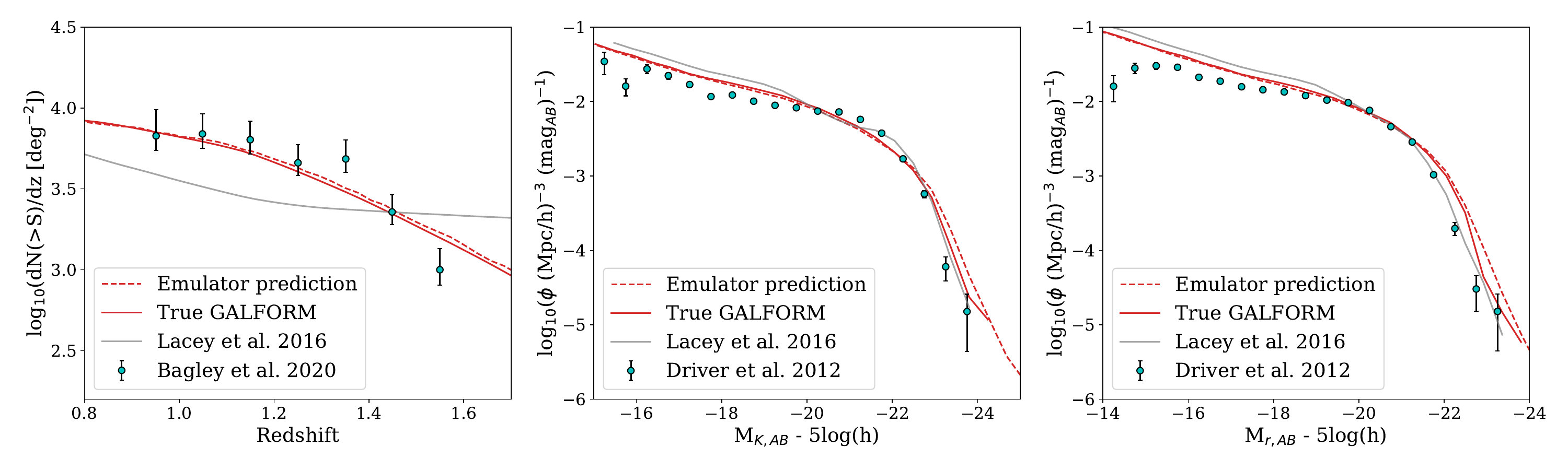}
    \caption{Predictions for the calibration data from the lowest MAE parameter set as evaluated by \texttt{GALFORM} (solid red) compared with the equivalent parameters evaluated by our emulator (red dashed) with the calibration data described in \S\ref{sec:CalibrationData}. The grey line shows the predictions of the \citet{Lacey2016} model.}
    \label{fig:BestMCMC}
\end{figure*}

Out of the lowest MAE parameters from the 100 MCMC chains, we plot the output from the 50 best sets of parameters evaluated using \texttt{GALFORM}  in Fig.~\ref{fig:50BestMCMC}. These runs cover a range of weighted MAE, from $0.25-0.31$, with the remaining runs extending to a weighted MAE of 0.64. 
The 50 best-fitting runs characterise the constraint datasets well and confirm the effectiveness of our MCMC optimisation and emulator while also indicating the level of uncertainty present in our method. We show the run with the lowest MAE in Fig.~\ref{fig:BestMCMC}, along with the emulator prediction for the same set of parameters (red dashed line), along with the output of the model presented in \citet{Lacey2016} as the solid grey line. We see that there is a spread of possible parameters. Therefore the best-fitting parameters presented are just one realization of many possible choices due to the effects of calibrating to multiple data sets with tensions between them and the degeneracies between the parameters.

The spread across the 50 best MCMC chains as evaluated by \texttt{GALFORM} is tight across the $K$- and $r$-band LFs; there is somewhat more variance in the redshift distribution outputs particularly at higher redshifts. The redshift distribution predicted using our overall best-fitting set of parameters is within the error bars of most of the \citet{Bagley2020} data points. Due to the tension between the H$\alpha$  redshift distribution and the local LFs, the general trend of fits to the LFs is to over-predict the bright end. This is particularly evident for the $z=0$ $r$-band LF, although the selected parameters do well at replicating the break. There is greater uncertainty in the fits to the $z=0$ $K$-band LF. The lowest weighted MAE parameter set predicts a far weaker break compared to the data from \citet{Driver2012}. In Table~\ref{tab: Bestrange} we show the set of parameters with the lowest weighted MAE to the observational data (corresponding to the red line in Figs.~\ref{fig:50BestMCMC} and \ref{fig:BestMCMC}), 
and compare with the parameters presented in \citet{Lacey2016} (hereafter named Lacey16). 
We also show the parameter set, which, out of the 50 best-fitting models, is the closest in parameter space to the Lacey16 model. 
Looking at this model and the parameter values from the 50 best MCMC chains in general, 
we find that certain parameters, such as $V_{\textrm{SN,disk}}$ and $\gamma_{\textrm{SN}}$ are constrained to a tight range of values, whereas parameters such as $V_{\textrm{SN,burst}}$, $f_{\textrm{burst}}$, and $\tau_{*\textrm{burst,min}}$ can be drawn from a large proportion of the explored range. Although we do see some parameters that have a large proportion of their parameter spaces sampled, the distribution is not always uniform. The $\nu_{\textrm{SF}}$ parameter appears to cover a very large range. However, when looking at the corner plot of Fig.~\ref{fig:Corner} we see that the majority of the sampling occurs at the high values of $\nu_{\textrm{SF}}$ but there is a small sub-region sampled at $\sim1.0$. 
The parameter $f{_\textrm{SMBH}}$ sampling distribution is skewed left which extends the accepted parameter range.

\begin{table}
 \caption{Results from the 50 best-fitting MCMC chain parameters (as measured by the weighted MAE in Eqn.~\ref{eq: MAEmod}) found using the emulator. The first column gives the parameter name. In the second column, we present the parameters for the best fit seen in Fig.~\ref{fig:BestMCMC}.
 The \citet{Lacey2016} model parameters are given in the third column. The final column gives the parameters of the model drawn from the 50 best-fitting models whose parameters are closest to those in Lacey et~al.}
 \label{tab:paramrange}
 \begin{tabular*}{\columnwidth}{
 @{}
 l@{\hspace*{17pt}}
 l@{\hspace*{17pt}}
 l@{\hspace*{17pt}}l@{\hspace*{17pt}}
 }
  \hline
  Parameter & This work & Lacey16 &  Lacey16-like example\\
  \hline
   $\nu_{\textrm{SF}}$ & 3.97 & 0.74 & 2.81\\ 
   $V_{\textrm{SN, disk}}$ [kms$^{-1}$] & 201.30 & 320 & 248.21\\
  $V_{\textrm{SN, burst}}$ [kms$^{-1}$] & 785.64 & 320 & 765.76\\
  $\gamma_{\textrm{SN}}$ & 3.98 & 3.40 & 3.98 \\
    $\alpha_{\textrm{ret}}$ & 0.27 & 1.00 & 1.08\\
  $\textrm{F}_{\textrm{stab}}$ & 0.85 & 0.90 & 0.85 \\
  $f_{\textrm{ellip}}$ & 0.22 & 0.30 & 0.04\\
  $f_{\textrm{burst}}$ & 0.083 & 0.05 & 0.05\\
  $\tau_{\textrm{* burst,min}}$ [Gyr] & 0.032 & 0.10 & 0.11\\ 
    $f_{\textrm{SMBH}}$ & 0.039 & 0.005 & 0.05\\
  $\alpha_{\textrm{cool}}$ & 0.79 & 0.80 & 0.99\\
  \hline
  \label{tab: Bestrange}
 \end{tabular*}
\end{table}

We compare the weighted MAE of our best-fitting model with the Lacey16 model, using the procedure described in \S\ref{sec:rParameter}, that is using the same weighting scheme we have been using up to this point. Using this metric, as expected, the new model is a better overall fit to the calibration data, with a weighted MAE of 0.25, compared with 0.50 for Lacey16. The MAE for Lacey16 is outside the range of the minimum MAE reached by our 50 best MCMC chains but within the range of lowest MAE values from the 100 MCMC chains. The improved MAE of the new best-fitting model (and indeed the majority of our MCMC-found models) is mainly due to the large improvement in the fits to the H$\alpha$ redshift distribution, while the fits of the new models to the $K$- and $r$-band LFs are similar to those of the Lacey16 model. The new model fit is closer to the faint end of the observed LFs, whereas the Lacey16 model is closer to the \citet{Driver2012} datapoints at the bright end, particularly in the $r$-band. The main source of error for the Lacey16 model is the poor fit to the H$\alpha$ redshift distribution, whereas our model more accurately describes the drop off in number counts beyond $z\sim1.4$. This can be quantified by considering the contribution to the MAE from each statistic: the best-fitting model has an MAE of 0.09 for the H$\alpha$ redshift distribution whereas the Lacey16 model is worse with a MAE of 0.26. 
Although by eye the fits to the $K$-band luminosity functions are similar between the new model and Lacey16, the MAE values indicate that the new model fits better to the \citet{Driver2012} data than the Lacey16 model does: the new model has an MAE of 0.17, whereas the Lacey16 model achieves 0.20. This is likely to be due to closer fits at the faint end contributing to a higher proportion of the MAE score. The fit to the bright end is very similar but the Lacey16 model has a much sharper break in its luminosity function. We can break down the $K$-band LF MAE calculation further by focusing on the bright half of the observed data points. As mentioned above the Lacey16 model is a closer fit to the data points at the bright end as measured by eye. This is confirmed as the MAE of the bright part for the Lacey16 model is 0.11 and for our new model is 0.14. Finally, our model performs slightly better than the Lacey16 model when predicting the $r$-band LF, scoring an MAE of 0.23 vs 0.25 for the Lacey16 model. This is likely to be for similar reasons as the $K$-band, where our model is closer to the \citet{Driver2012} data at the faint end. It is clear that the Lacey16 model is slightly better at predicting the luminosity function from the exponential break to the bright end as our model over-predicts the bright end.
If we focus only on the bright half of the luminosity function, the Lacey16 model is a closer fit to the observed data than our model, with an MAE of 0.10 vs 0.16. 

Due to the tensions between the calibration data, better fits to the H$\alpha$ redshift distribution data come at the expense of more severe over-predictions ogf the bright-end of the LFs as previously discussed, and as shown by the lines when increasing the weighting in Fig.~\ref{fig:weightingtest}. Similarly, if we try to improve the fits to the LFs, this leads to an overestimation of the number of H$\alpha$ emitters at higher redshifts. 

\subsubsection{Number count predictions for the \textit{Euclid} redshift survey}
\label{sec:numbercounts}

\begin{figure}
    \centering
    \includegraphics[trim=0.5cm 1.3cm 0 0.5cm, width=1.01\columnwidth]{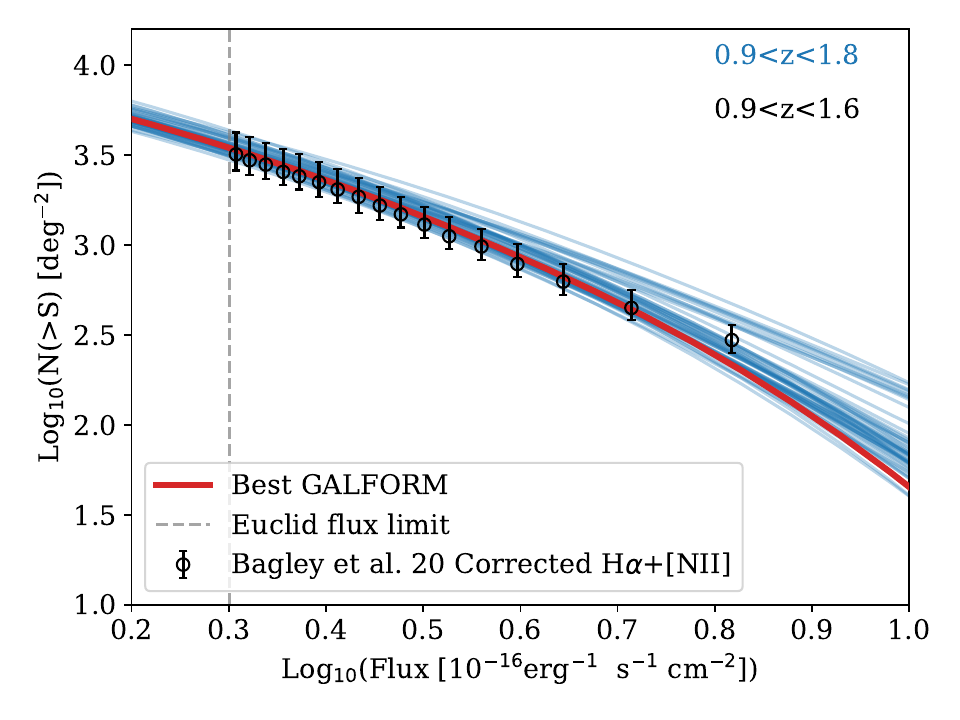}
    \caption{Number counts predictions from our 50 best MCMC parameters for galaxies with $0.9<z<1.8$ (blue lines), with the best set of parameters as evaluated by \texttt{GALFORM} in red. We plot this against the \citet{Bagley2020} $0.9<z<1.6$ number counts (black points). The Euclid flux limit $2\times10^{-16}~\textrm{erg}^{-1}~\textrm{s}^{-1}~\textrm{cm}^{-2}$ is marked by the vertical dashed grey line.}
    \label{fig:numbercounts}
\end{figure}

Galaxies detected through their emission in the unresolved   H$\alpha$(+N[II]) lines are the main target for the  \textit{Euclid} and \textit{Roman} redshift surveys.  Satisfied with the best fitting parameters from the MCMC search using the emulator as evaluated using \texttt{GALFORM}, we can use these models to predict the number of galaxies that will be seen by the upcoming surveys. The cumulative number counts are shown in Fig.~\ref{fig:numbercounts}, along with the recent WISP+3D-HST data from \citet{Bagley2020} for galaxies in the redshift range $0.9\le z \le1.6$. The corrected number counts from \cite{Bagley2020} for the WISP+3D-HST data at the \textit{Euclid} flux limit is $3266\substack{+157.7 \\ -174.8}\,\,$ H$\alpha$+N[II] emitters deg$^{-2}$. 
Our models predict the galaxy density in the slightly broader redshift range $0.9<z<1.8$, which matches that of the \textit{Euclid} redshift survey. 
From the 50 best models, the spread in emission-line number counts estimates for galaxies with a flux greater than the \textit{Euclid} limit ($f\ge2\times10^{-16}~\textrm{erg}^{-1}~\textrm{s}^{-1}~\textrm{cm}^{-2}$) is 2962-4331 deg$^{-2}$, with our best-fitting  model to the calibration data outputting a number count of 3462.5 deg$^{-2}$, corresponding to $\sim46.7$ million sources over 13,500 deg$^2$. Our best-fitting model comfortably lies within the range of the \citet{Bagley2020} H$\alpha$+N[II] number counts. The distribution of predicted number counts can also be quantified using the 10-90 percentile range of the 50 best models, which gives the narrower spread of  3158-3952 deg$^{-2}$. 
We compare our number count predictions with those of \citet{Pozzetti2016} who empirically fit luminosity functions to earlier surveys, HiZELS, WISP and HST+NICMOS. Covering the redshift range $0.9<z<1.8$ to a flux limit of $2\times10^{-16}~\textrm{erg}^{-1}~\textrm{s}^{-1}~\textrm{cm}^{-2}$, Pozzetti et~al. predicted 2000-4800 H$\alpha$ emitters deg$^{-2}$.
 It is worth noting that the \citet{Pozzetti2016} predictions are in terms of observed H$\alpha$ flux, i.e. they 
 are corrected for [NII] contamination. In contrast, our results blend H$\alpha$+N[II] to match the results of \citet{Bagley2020}. At the spectral resolution of \textit{Euclid}, these two lines will be partially blended.

Fig~7 of \citet{Bagley2020} shows the observed cumulative number counts along with fits from various models including the three empirical models from \citet{Pozzetti2016}. For the purposes of this comparison, \citet{Bagley2020} converted the H$\alpha$ counts from the \citet{Pozzetti2016} models to H$\alpha$+[NII] counts using a fixed [NII]/H$\alpha$ line ratio: H$\alpha$=0.71 (H$\alpha$+[NII]).
Out of their three models, the only one that fits the $0.9\le z \le1.6$ observations well is Model 3 which shows a similar fit to our best fitting model in Fig.~\ref{fig:numbercounts}. Fig.7 from \cite{Bagley2020} also shows the redshift distribution predictions from \citet{Pozzetti2016}, where once again Model 3 is the best fit to the observed redshift distribution. However, the fits are only good for the first five data points before the drop off in counts observed for $z\sim1.4$. As seen in Fig.~\ref{fig:BestMCMC}, our best-fitting model is a better fit to the observed redshift distribution data points as we represent more closely the trends beyond $z\sim1.4$.

We also calculate the number counts for a \textit{Euclid}-like survey with a magnitude limit of $H=24$ using our best model, keeping the line flux limit and redshift range fixed. The number of galaxies counted in this case is 3444.5 deg$^{-2}$; including the $H$-band cut reduces this by $0.5$ per cent, which is somewhat smaller than the 3 per cent reduction reported by \cite{Zhai2021}. 

\subsubsection{Number count predictions for \textit{Roman}}
\label{sec:numbercountsWFIRST}

\begin{figure}
    \centering
    \includegraphics[trim=0.5cm 1.3cm 0 0.5cm, width=1.01\columnwidth]{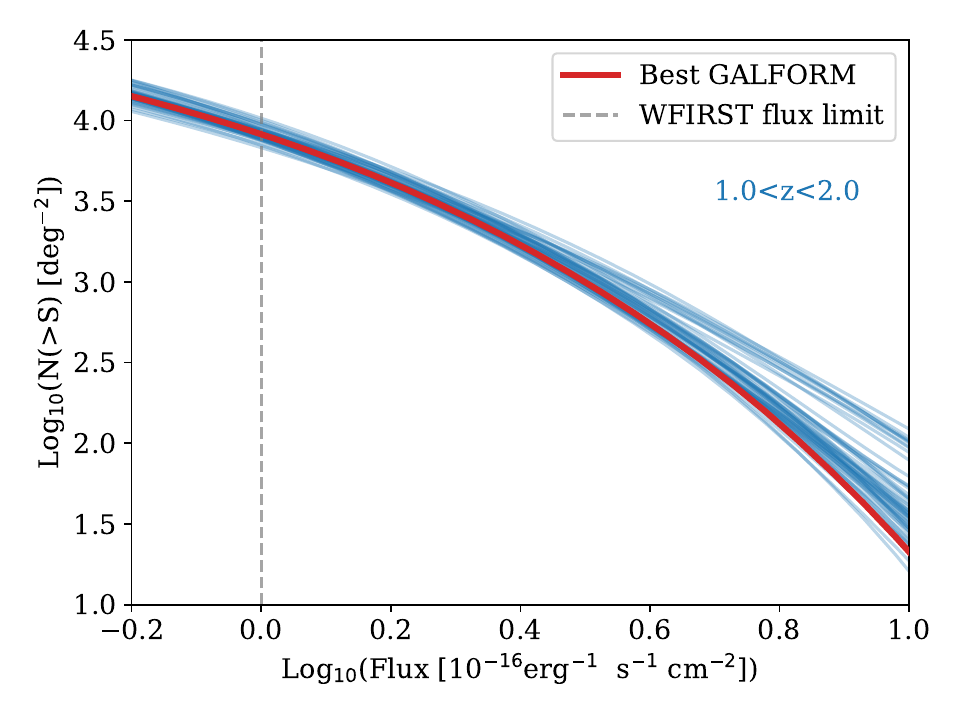}
    \caption{\textit{Roman} number counts predictions from our 50 best MCMC parameters for galaxies between $1.0<z<2.0$ (blue lines), with the best set of parameters as evaluated by \texttt{GALFORM} in red. The \textit{Roman} flux limit  $1\times10^{-16}~\textrm{erg}^{-1}~\textrm{s}^{-1}~\textrm{cm}^{-2}$ is shown by the vertical dashed grey line.}
    \label{fig:numbercountsWFIRST}
\end{figure}

The High Latitude Spectroscopic Survey onboard NASA's \textit{Nancy Grace Roman Space Telescope} will cover 2000 deg$^2$ and will use H$\alpha$(+N[II]) galaxy redshifts to map large-scale structure at $1<z<2$ \citep{Spergel2015} to a flux limit of $1\times10^{-16}~\textrm{erg}^{-1}~\textrm{s}^{-1}~\textrm{cm}^{-2}$. We use the same 50 best-fitting parameters described in \S\ref{sec:rParameter} to evaluate \texttt{GALFORM} to predict the number of galaxies that will be seen by a \textit{Roman}-like survey.
The cumulative number counts are shown in Fig.~\ref{fig:numbercountsWFIRST}. From the  50 best models, the spread in number counts estimates for H$\alpha$ sources seen by \textit{Roman} is 6786-10322 deg$^{-1}$, with the same best model as described in \S\ref{sec:rParameter} outputting a number count of 8212.5 deg$^{-1}$. This corresponds to $\sim$16.4 million sources over the 2000 deg$^2$ survey. Our best-fitting model agrees with the number counts predicted by \citet{Zhai2019} who used  \texttt{GALACTICUS}. 
The 10-90 percentile range of the counts is 7536-9470 deg$^{-1}$. 

\subsubsection{Predictions for the evolution of galaxy bias}
\label{sec:bias}

As our model is physically motivated and connects galaxies to dark matter halos, 
we can also use \texttt{GALFORM} to predict the effective clustering bias as a function of redshift.  The bias is a direct input into the calculation of the signal-to-noise of the clustering measurements. We calculate the asymptotic effective bias in real space using the \texttt{COLOSSUS} package \citep{Diemer2018}, choosing the numerically calibrated bias - halo mass model from \citet{Tinker2010}.

\begin{figure*}
    \centering
    \includegraphics[trim=1.5cm 1.cm 1.0cm 0.7cm, width=1.95\columnwidth]{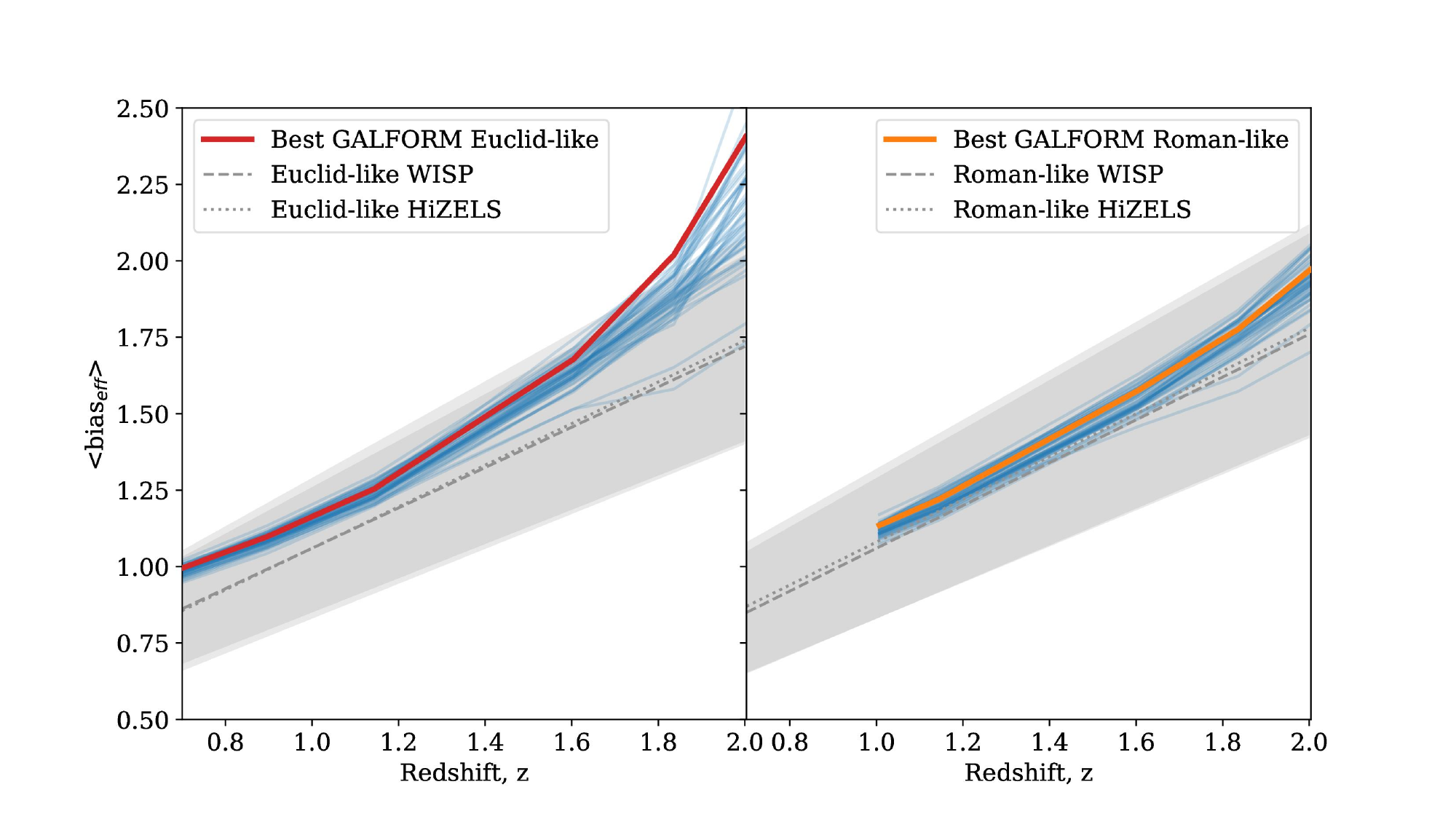}
    \caption{
    The effective clustering bias for a \textit{Euclid} limited survey (left) and \textit{Roman} (right). In both cases, the blue curves show the 50 best models as a function of redshift, evaluated using \texttt{GALFORM} and using the Colossus routines for computing bias as a function of host halo mass. We have highlighted our best-fitting model to the calibration dataset as red (left) or oragne(right) lines. We also plot the fits to the bias predictions from \citet{Merson2019} when adopting a WISP-calibrated lightcone (grey dashed line) and a HiZELS-calibrated lightcone (grey dotted line) and their uncertainty (shading).
    }
    \label{fig:bias_WFIRST}
\end{figure*}

In Fig.~\ref{fig:bias_WFIRST} we plot the results for the effective linear bias ($b_{\textrm{eff}}$) in real-space as a function of redshift for a \textit{Euclid}-like survey (left panel) and for a \textit{Roman}-like survey (right panel). We 
compute the effective bias for halos containing an H$\alpha$ emitter that is brighter than the flux limit of the corresponding survey and in the expected redshift range. 
We find similar results for the linear real-space bias for the \textit{Euclid} and \textit{Roman} selections. At lower redshifts, the predicted bias has a linear dependence on redshift. This steepens at the highest redshifts shown. 
The dashed and dotted grey lines show the \citet{Merson2019} models of the linear bias evolution with a WISP- and HiZELS-calibrated models respectively. Their results show a linear dependence of the effective bias on redshift. 
 Merson~et~al calibrate their dust extinction specifically to reproduce the observed H$\alpha$ luminosity functions. 
 Finally, differences in the choice of cosmologies 
 and bias-halo mass relations will cause slight discrepancies between our predictions and those of Merson et~al. 



\subsubsection{Comparison to older calibration datasets}
\label{sec:Compareextended}

\begin{figure}
    \centering
    \includegraphics[trim=0.5cm 1.1cm 0 0.5cm, width=0.98\columnwidth]{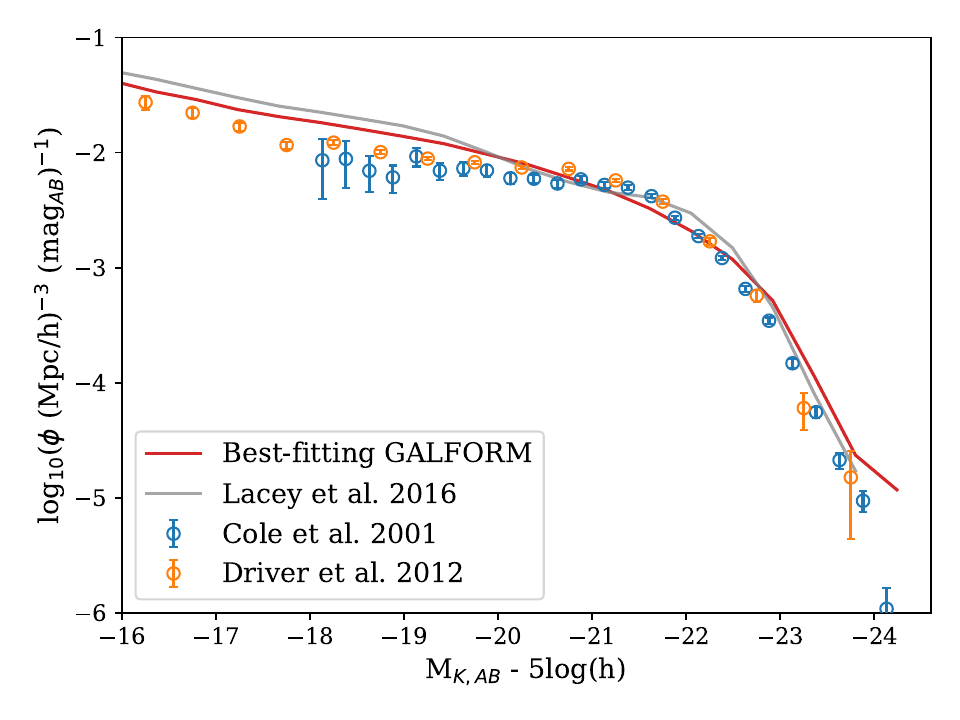}
    \caption{The prediction of the new \texttt{GALFORM} variant (red line) for the $z=0$ $K$-band LF compared with the \citet{Driver2012} (orange) and  \citet{Cole2001} (blue) data sets. We also plot the \texttt{GALFORM} model by \citet{Lacey2016} (grey line), which was calibrated by hand to several datasets including the Cole et~al. LF. Here, we calibrate to the \citet{Driver2012} data. }
    \label{fig:BestvsCole}
\end{figure}

Our best-fitting model is calibrated to the local $K$- and $r$-band LFs from \cite{Driver2012}. Therefore we have expanded the comparison datasets to include an older $K$-band LF from \citet{Cole2001} which was used in the calibration of many previous \texttt{GALFORM} variants. In Fig.~\ref{fig:BestvsCole} we plot our best-fitting \texttt{GALFORM} model $z=0$ $K$-band LF, found using our emulator-based MCMC calibrated to the \citet{Driver2012} LF, and compare this with the \citet{Cole2001} $K$-band LF data. We also plot the \citet{Driver2012} $K$-band LF for comparison. We see that the \citet{Cole2001} and \citet{Driver2012} data agree reasonably well, particularly for bright galaxies. The consistency between the new local calibration data and the old calibration data indicates that the two observational LFs agree with one another. Therefore, our \texttt{GALFORM} prediction agrees as well with the \citet{Cole2001} data as it does for the \citet{Driver2012} data, up to faint galaxies where the \citet{Cole2001} data is noisier. The \citet{Cole2001} LF estimate overlaps mainly with the brighter \citet{Driver2012} data (as expected given the greater depth of the GAMA survey compared with the 2dFGRS and 2MASS data used by Cole et~al.), and as we have seen in our previous analyses, the weighting scheme compromises our new model at the bright end. Therefore, the new model achieves poorer fits at the bright end when compared to the Lacey16 model for the \citet{Cole2001} calibration data also. Our model across all data points scored an MAE of 0.27 compared to the \citet{Cole2001} data; this is worse than the Lacey16 model which achieves an MAE of 0.23. Our model scores worse for similar reasons as previous analysis on the \citet{Driver2012} data; over-prediction at the very bright end, and a much shallower turn-over. Nevertheless, our model is still a good approximation to the \citet{Cole2001} data.

\section{Discussion and Conclusions}
\label{sec:Discussion}

We have presented a method for efficiently exploring and calibrating the \texttt{GALFORM} semi-analytical galaxy formation model across a wide range of outputs, building on \cite{Elliott2021}. Whereas Elliott et~al. focused on using different local datasets in their model calibration, we have also used data at intermediate redshifts, specifically to find models that reproduce current data on the redshift distribution of H$\alpha$ emitters. We calibrated the model over an eleven-dimensional subset of the full model parameter space. We used a deep learning method to mimic running the full \texttt{GALFORM} model. Training the emulator required of the order of 1000 full model runs. With the emulator, we explored the parameter space using MCMC walkers.  


We calibrated the model to three sets of observational data: the $z=0$ galaxy LFs in the $r$- and $K$-bands from \cite{Driver2012} and the redshift distribution of H$\alpha$ emitters at intermediate redshifts from \cite{Bagley2020}. However, we did not consider the observational error bars during the model exploration. Instead, we used an absolute error metric (MAE) to quantify the distance between the emulator output and the full model calculations. Hence, it is difficult to provide meaningful uncertainties on the best-fitting parameters. We give an illustration of the uncertainty on the model predictions by plotting the results from the best-fitting model for each MCMC walker, as judged by the model that returned the smallest MAE. 
We have discovered tensions between the calibration datasets and the model predictions as we could not find equally good fits when all data sets are weighted equally in the MCMC search. The weight given to the H$\alpha$ redshift distribution constraint was increased, moving to a different region of parameter space which modified the fits to the $K$- and $r$-band LFs, leading to over-prediction at the bright end.

Similarly, we have not considered the uncertainties associated with the emulator. 
There are two types of uncertainty to account for when emulating model outputs: the uncertainty due to the emulator parameters (that is the weights of the neural network), and the uncertainty inherent in the data generation process (for example, the sampling noise on the \texttt{GALFORM} outputs, such as the bright end of the LF where there are few galaxies). The network hyperparameter space was explored using a trial-and-error process to justify the choice of network architecture. We further reduce uncertainties relating to the weights of the emulator by ensembling individual network estimates.

The majority of variance in the output of our model is due to a few key parameters, which leads to tensions when trying to calibrate to multiple observational datasets. The tensions between the observed datasets were explored, using our MCMC algorithm to fit the emulator output to the constraints, eventually finding the weighting scheme for a global fit to the observations. With this, we find a set of parameters which provides an improved fit to the redshift distribution data as compared with an earlier version of a \texttt{GALFORM} model presented in \citet{Lacey2016}. We go further by producing number count predictions for a \textit{Euclid}-like survey using our best model, improving on previous empirical models by \citet{Pozzetti2016} by using more recent and complete datasets from \citet{Bagley2020}. For a flux limit of $2\times10^{-16}~\textrm{erg}^{-1}~\textrm{s}^{-1}~\textrm{cm}^{-2}$ between the redshift range $0.9<z<1.8$, our 50 best models predict 2962-4331 H$\alpha$ emission-line sources deg$^{-2}$, with 3158-3952 sources deg$^{-2}$ between the 10th and 90th percentile. Our best-fitting model estimates 3462.5 sources deg$^{-2}$, which is comparable to the \citet{Bagley2020} observation. The predictions we produce for the number of galaxies estimated to be seen from the Euclid wide field are more constrained than previous models and are better in line with the recent observed number counts. Adding a requirement that the sources are also brighter than $H=24$ removes only 0.5 per cent of the emitters. 

As we are using a physical model that connects galaxies to their host dark matter halos, we can predict the clustering of H$\alpha$ emitters. Our bias predictions are similar to those of \citet{Merson2019}, but with some differences in detail: Merson et al. found that their bias prediction has a linear dependence on redshift, whereas we find that the bias evolves somewhat more rapidly at higher redshift.  
  We also produce number count and bias predictions for the upcoming \textit{Nancy Grace Roman} mission which will survey similar sources to \textit{Euclid} but over a slightly different redshift range and to a deeper flux. 
Similar results to ours, but without the extensive parameter search and emulation of the semi-analytic model have been presented by \cite{Zhai2019,Zhai2021,Wang2022}.



We have shown that the method used by \cite{Elliott2021} to automate the calibration of \texttt{GALFORM} can be applied to calibration data that include intermediate redshift observations. Elliot et~al. (2024, in preparation) address a similar data calibration challenge using an even more efficient method, Bayesian optimisation. 

\vspace{-0.5cm}

\section*{Acknowledgements}
We acknowledge comments from Cedric Lacey. 
MM was supported by a CASE PhD Studentship from the Science and Technology Facilities Council (STFC ST/S005617/1). CMB and DS acknowledge support from STFC (ST/T000244/1, ST/X001075/1). We used the DiRAC@Durham facility managed by the Institute for Computational Cosmology on behalf of the STFC DiRAC HPC Facility (www.dirac.ac.uk). The equipment was funded by BEIS capital funding via STFC grants ST/K00042X/1, ST/P002293/1, ST/R002371/1 and ST/S002502/1, Durham University and STFC operations grant ST/R000832/1. DiRAC is part of the National e-Infrastructure.


\vspace{-0.75cm}

\section*{Data Availability}
The \texttt{GALFORM} outputs, parameter values, and best-fitting parameters may be shared on reasonable request to the corresponding author.



\vspace{-0.5cm}

\bibliographystyle{mnras}
\bibliography{main} 





\bsp	
\label{lastpage}
\end{document}